\documentclass[12pt,aps,showpacs,nofootinbib,oneside,hidelinks,a4paper,english]{revtex4-2}

\usepackage{graphicx}
\usepackage{amssymb}
\usepackage{amsmath}
\usepackage{graphics}
\usepackage{dcolumn}
\usepackage{bm}
\usepackage{color}
\usepackage{epstopdf}
\usepackage{lipsum}
\usepackage{ulem}
\begin{document}

%\preprint{APS/123-QED}

%\title{Non-Hermitian Deformations of Inflation with a Real Background Evolution}
%\title{Complex Inflaton Potentials and Effective Reheating: A Robust Framework Consistent with Planck Data}
\title{Complex Inflaton Potentials with Nonminimal Coupling: Robust Inflation and Geometric Reheating}

\author{S. D.  Campos}\email{sergiodc@ufscar.br}
	
\affiliation{Applied Mathematics Laboratory-CCTS/DFQM, Federal University of São Carlos, Sorocaba, CEP 18052780, Brazil}
        
\begin{abstract}
We investigate an inflationary scenario driven by a complex scalar field nonminimally coupled to gravity and subject to a non-symmetric complex potential. The real part of the potential controls the cosmological background and realizes a plateau-type inflation compatible with $\alpha$-attractor $\mathrm{T}$-models, while the imaginary part acts as an effective non-Hermitian deformation encoding dissipative effects. Working in the Jordan frame and imposing ghost-free conditions on the effective Planck mass, we derive the background equations and define a complex equation-of-state parameter whose real part governs the expansion and whose imaginary part quantifies departures from conservative dynamics. Numerical integration shows that the duration of inflation is primarily controlled by the nonminimal coupling $\zeta$, whereas the complex asymmetry parameter $\Delta\varepsilon$ has a negligible impact on the real background: the real energy density and pressure vary by less than $10^{-5}$ as $\Delta\varepsilon$ is scanned over its allowed range. Mapping the two-field dynamics to an effective single-field description in the Einstein frame, we obtain a spectral index 
$n_s\simeq 0.968–0.971$ and a tensor-to-scalar ratio 
$r<10^{-3}$, fully consistent with Planck 2018 bounds. We introduce a relevance parameter and show that non-Hermitian effects remain strongly suppressed during slow roll but grow to 
$\mathcal{O}(1)$ near the end of inflation, triggering an efficient reheating phase without additional fields or {\it ad hoc} friction terms. In this sense, the imaginary sector behaves as an effective $\mathcal{PT}$-symmetric channel for energy transfer, providing a geometrical mechanism for inflation and its exit within a non-Hermitian scalar–tensor framework.
%We investigate inflation driven by a complex scalar field with a non-symmetric complex potential and non-minimal coupling to gravity. This framework naturally splits the dynamics into a real sector (which controls the cosmological background) and an imaginary sector (which governs dissipative effects). Numerical integration shows that the duration of inflation is controlled primarily by the non-minimal coupling $\zeta$, whereas the complex asymmetry parameter $\Delta\varepsilon$ has a negligible impact on the background evolution. This robustness is confirmed through sensitivity analysis: the real energy density and pressure vary by less than $10^{-5}$ as $\Delta\varepsilon$ is scanned across the parameter space. Observationally, the present model produces a spectral index $n_s \simeq 0.968$--$0.971$ and a tensor-to-scalar ratio $r < 10^{-3}$, consistent with Planck 2018 constraints. The observables remain insensitive to $\Delta\varepsilon$: predictions from different non-Hermitian strengths overlap in the $n_s-r$ plane, demonstrating that complex deformations of the inflaton potential do not compromise agreement with the data. Moreover, non-Hermitian effects become dynamically active near the end of inflation, where the relevance parameter grows to $\mathcal{R} \sim \mathcal{O}(1)$ and oscillates chaotically. This behavior provides a natural, geometrically motivated mechanism for effective reheating without introducing extra fields or \textit{ad hoc} friction terms.
\end{abstract}

\date{\today}

\maketitle

%%%%%%%%%%%%%%%%%%%%%%%%%%%%%%%%%%%%%%%%%%%%%%%
\section{Introduction}\label{sec:intro}

We know the Universe has been accelerating for almost the last 30 years \cite{riess.1998,perlmutter.1999}, and since then, our understanding of cosmic evolution has undergone substantial refinement. However, this does not mean that we have solved the problem of defining the physical entity that drives inflation itself and its exit. Inflationary cosmology, traditionally formulated using real scalar fields with Hermitian potentials, ensures real energy densities and unitary dynamics \cite{jaman_sami.2022}. This framework successfully explains the flatness and homogeneity of the Universe but is not unique within effective field theory (EFT) \cite{berera.1995,kinney.2005,cheung.2008,kobayashi.2011,martin.2014,baumann.2015,pinol.2015,achucarro.2018,cabass.2021}. In particular, nothing in EFT forbids complex or asymmetric inflaton potentials \cite{yurov.2002,dias.frazer.2016,zhang.zheng.2022}, especially if the inflaton sector behaves as an open system interacting with gravity \cite{berera.1995,basterogil.2021}.

Complex scalar fields appear in quantum field theory (QFT), where complex masses describe instability and dissipation \cite{denner.2020,ashida.2020}. In cosmology, such terms are rarely explored explicitly at the potential level. When included, they lead to non-Hermitian dynamics and possible instabilities that must be analyzed to distinguish non-physical ghosts from meaningful effects, such as tachyonic growth. As shown by Chavanis \cite{chavanis.2022}, complex fields can still yield consistent large-scale cosmological behavior. Furthermore, as shall be seen, the use of scalar fields for inflation provides a description in which reheating and decay arise without the need for extra fields or \textit{ad hoc} friction terms.

From an EFT perspective, non-Hermitian effects mediated by complex fields or by dissipation could have influenced inflation or its end. Such extensions provide a phenomenological way to model irreversible dynamics and symmetry breaking without specifying a microscopic baryogenesis mechanism \cite{ahmad.2019}. The main goal of this work is to demonstrate that a complex inflaton potential generates the non-equilibrium conditions and effective symmetry breaking required for baryogenesis. However, calculating the precise baryon asymmetry of the Universe would require specifying the explicit couplings between the inflaton and Standard Model fields (e.g., via a specific reheating temperature or sphaleron process), which introduces model-dependent parameters unrelated to the core inflationary mechanism. Therefore, we restrict our discussion of baryogenesis to a qualitative level, establishing that the necessary dynamical ingredients are present without committing to a detailed particle physics realization.

Specifically, we analyze an inflationary model with a complex inflaton field (CIF) \cite{khalatnikov.1992,khalatnikiv.1994,yurov.2002} subject to a non-symmetric complex potential. Independent couplings for the field and its conjugate define a generally complex mass matrix, which can be exactly diagonalized to yield normal modes with real or complex squared masses \cite{denner.2020,yurov.2002}. This procedure preserves a positive kinetic sector, avoiding ghost instabilities while allowing for physically interpretable growth or decay modes. 

We also assume a non-minimal coupling to gravity that leads to energy exchange between the scalar field and the non-minimal gravity sector \cite{bertolami.2008,faraoni.2004}. In the numerical analysis, the background evolution is obtained by integrating the Jordan-frame equations, while the Einstein frame is used only to interpret the resulting dynamics and conservation laws. 

The framework of this manuscript addresses several questions: Can complex potentials sustain real, stable inflationary backgrounds? Under what conditions do imaginary potential terms remain small during slow roll but affect late-time dynamics? Can complex structures account for reheating without extra fields? How do non-minimal coupling and complex asymmetry control the duration of inflation and non-conservative effects? As shall be seen, the numerical investigation provides answers to all these questions.  

The paper is organized as follows. Section \ref{sec:section2} introduces the model and analyzes stability and 
ghost conditions. In Section \ref{sec:lagrangian}, we couple the fields to gravity and derive the equations of motion in 
the Jordan frame. In Section \ref{sec:eos}, we define the dynamical equation of motion and introduce probes for 
non-Hermitian effects. A description of the effective reheating from non-Hermitian effects is also presented. Section \ref{sec:comp} compares the model with standard reheating scenarios and presents comparisons with Planck data. Finally, in Section \ref{sec:final}, we present remarks and discuss future research.

%%%%%%%%%%%%%%%%%%%%%%%%%%%%%%%%%%%%%%%%%%%%%%%
\section{Complex Inflaton Field and Complex Potential Energy}\label{sec:section2}

\subsection{Complex Inflaton Field}

As noted long ago, complex fields emerge in modern theories of particle physics \cite{I.M.Khalatnikov.1997}, and it seems natural to extend this mathematical tool to inflationary models. As is well known, a complex scalar field offers some technical and conceptual advantages over a purely real one. First, it admits the most general quadratic structure in EFT \cite{burgess.2007}, where the field and its complex conjugate can have independent couplings, yielding asymmetric, generally complex mass matrices that a single real scalar cannot reproduce. Second, the CIF possesses a conservative dynamical sector determined by the real part of the potential, alongside non-conservative effects in its imaginary part, which are useful for studying instabilities, decay, and energy transfer in an expanding Universe \cite{burgess.2007,denner.2020}. At the phenomenological level, we assume that the imaginary sector works as a thermal reservoir, thereby enabling a coarse-grained description of the reheating process. A fully specified microphysical framework, however, would be required to determine the ultimate channels of energy transfer and dissipation, for instance, by explicitly introducing couplings to light fermionic degrees of freedom.

The massive inflaton field $\Phi$ is assumed to have a nonzero classical expectation value and can be written as a complex function of two real scalar fields, $\phi$ and $\chi$ (hereafter, $c=G=\hbar =1$). Thus, one defines $\Phi$ as
\begin{eqnarray}
\Phi=\frac{1}{\sqrt{2}}(\phi+i\chi), ~~ \Phi^{*}=\frac{1}{\sqrt{2}}(\phi-i\chi),
\end{eqnarray}
where $\phi=\phi(t)$ and $\chi=\chi(t)$, result in
\begin{eqnarray}
\nonumber \Phi\Phi^{*}=|\Phi|^2=\frac{1}{2}\bigl(\phi^2+\chi^2\bigr).
\end{eqnarray}

From an EFT perspective, a CIF enables the symmetry parameterization of the scalar sector. Treating $\phi$ and $\chi$ as independent variables allows one to write a renormalizable, local potential without imposing Hermiticity on quadratic operators \emph{a priori} \cite{burgess.2007,denner.2020}. %This formulation ensures operator completeness at the quadratic level and accommodates asymmetric mass terms, instabilities, and decay channels \cite{burgess.2007,denner.2020}. %Separating the real and imaginary parts of the potential organizes the dynamics, isolating conservative contributions that control the inflationary background from non-conservative effects relevant beyond the slow-roll approximation.

%============
\subsection{The Complex Potential}

The complex potential $V(\Phi)$ is defined as
\begin{eqnarray}\label{eq:pot2}
    V(\Phi)=V_R(\Phi)+iV_I(\Phi),
\end{eqnarray}
where its real ($V_R(\Phi)$) and imaginary ($V_I(\Phi)$) parts are real scalar functions. Taking into account the $\alpha$-attractor T-models \cite{kallosh.2013_1,kallosh_2013_2,kallosh.2014,dimopoulos.2018,shojaee.2021}, one writes (for a generic scalar field $\sigma$)
\begin{eqnarray}\label{eq:pot_flat1}
    V(\sigma)=V_0\tanh^{2n}\left(\frac{\sigma}{M_P\sqrt{6}\alpha}\right),
\end{eqnarray}
where $\alpha$ (the attractor) and $n$ are parameters of the model. Moreover, $M_P\simeq 10^{19}$ GeV is the usual Planck mass. This potential is quite similar to the KKLT-inspired potential \cite{mishra.2025}
\begin{eqnarray}\label{eq:pot_kklt_1}
    V(\sigma)=V_0\left(\frac{M_P^{2n}+\sigma^{2n}}{1+(\sigma/\mu)^{2n}}\right).
\end{eqnarray}

To see this, one assumes $n=1$, and for small $y=\eta/(M_P\sqrt{6\alpha})<\!\!<1$, one writes
\begin{eqnarray}
\nonumber    \tanh^2(y)=\frac{\sinh^2(y)}{\cosh^(y)}=\frac{e^{2y}-1}{e^{2y}+1}\approx \frac{1}{1+e^{-2y}},
\end{eqnarray}

Then, for $y<\!\!<1$, one has the result
\begin{eqnarray}\label{eq:pot_flat2}
    V(\sigma)\approx V_0\left(\frac{1}{1+e^{\sigma/(M_P\sqrt{6}\alpha})}\right),
\end{eqnarray}
which is a good approximation for the KKLT-inspired potential \eqref{eq:pot_kklt_1} for $\sigma\ll 1$ (and $M_P=1$)
\begin{eqnarray}\label{eq:flatx}
    V(\sigma)=V_0\left(\frac{1}{1+(\sigma/\mu)^{2}}\right).
\end{eqnarray}

The main rule of \eqref{eq:pot_flat1} (or \eqref{eq:pot_flat2}) is described as a long, flat potential that takes into account the necessary slow-roll during $\sim 10^{50-60}\times$ the Hubble radius. This can be achieved when $V'\approx V''\approx 0$, resulting in
\begin{eqnarray}\label{eq:ev}
    \epsilon_V=\frac{M_P^2}{2}\left(\frac{V'}{V}\right)\ll 1,
\end{eqnarray}
and
\begin{eqnarray}\label{eq:etav}
\nonumber    |\eta_V|=M_P^2\left| \frac{V''}{V}\right|\ll 1,
\end{eqnarray}
with prime denoting the derivative with respect to $\phi$ ($\chi$).

Considering the above discussion, the radial function $P(x)$ is defined as a smoothed step function, similar to the T-models of inflation mentioned previously. Then, one writes
\begin{eqnarray}\label{eq:plateau}
    P(x)=\frac{1}{1+(x/\mu)^s}
\end{eqnarray}
acting as an EFT radial form factor that modulates the real part of the inflaton potential without introducing additional degrees of freedom. The parameter \( x \equiv \sqrt{\phi^{2} + \chi^{2}} \) quantifies the amplitude of the CIF, being invariant under real orthogonal transformations (rotations) in the \((\phi,\chi)\) plane. It can, therefore, be interpreted as the radial magnitude of the field in the corresponding internal field space. The dimensionless exponent $s$, on the other hand, controls the sharpness of the transition between the polynomial regime ($r<\!\!<\mu$) and the asymptotic plateau ($r>\!\!>\mu$). In typical applications, one takes $s=2$ for a smooth transition, and $s\gtrsim 6$ signifies a well-defined plateau. Henceforth, we use $s=2$ for a smooth transition.

The parameter $0<\mu$ is the radial scale of $P(x)$ and controls how early, in terms of field amplitude, the system ``enters'' the plateau inflation regime and for how long this plateau can be exploited during slow-roll. A recent comparison of $\alpha$-attractor models with the latest cosmological data indicates that $\alpha=0.0962_{-0.00047}^{+0.00046}$ \cite{bhattacharya.2023}. Then, a simple comparison between potential \eqref{eq:pot_flat2} and \eqref{eq:flatx} (taking $\sigma=\phi^2+\chi^2$) and retaining only the first term of the power series of the exponential yields
\begin{eqnarray}
 \nonumber   \mu^2=\frac{\sqrt{6}\alpha}{1+\frac{\sqrt{6}\alpha}{\phi^2+\chi^2}}.
\end{eqnarray}

Taking into account $\alpha=0.0962$ and small values of $\phi=\chi=0.1\sim 0.01$ at the end of inflation, one expects $\mu\approx 0.1\sim0.03$. 

Considering the CIF and the preceding discussion, one assumes a radial plateau expressed as
\begin{widetext}
\begin{eqnarray}
\nonumber    V_R(\phi,\chi)&=&V_0\left[\frac{1}{2}(\varepsilon_\phi+\varepsilon_\chi)(\phi^2-\chi^2)
-\frac{1}{2}m^2(\phi^2+\chi^2)
+\frac{\lambda}{4}(\phi^2+\chi^2)^2\right]P(x)+V_0\bigl[1-P(x)\bigr]=\\
\nonumber\label{eq:pot_real}&=& V_0\bigl\{V_{p}P(x)+\bigl[1-P(x)\bigr]\bigr\}
\end{eqnarray}
\end{widetext}
where $\varepsilon_\phi$, $\varepsilon_\chi$, and $\lambda$ are the interaction coupling constants, and $V_p$ is the polynomial part of the real part of the potential. Without loss of generality, we fix the mean slope parameter by setting $(\varepsilon_\phi + \varepsilon_\chi)/2 = 1$.

The corresponding terms of \(V_p\) are derived from the following results
\begin{eqnarray}
\nonumber \frac{}{}(\varepsilon_\phi+\varepsilon_\chi)\bigl[\Phi^2+(\Phi^*)^2\bigr] &=& \frac{1}{2}(\varepsilon_\phi+\varepsilon_\chi)(\phi^2-\chi^2),\\
\nonumber -m^2 \Phi\Phi^* &=& -\frac{1}{2}m^2(\phi^2+\chi^2),\\
\nonumber \lambda(\Phi\Phi^*)^2 &=&\frac{\lambda}{4}(\phi^2+\chi^2)^2.
\end{eqnarray}

The first term $\sim (\phi^2-\chi^2)$ furnishes an asymmetric mass matrix, while $\sim (\phi^2+\chi^2)$ represents the radial contribution. Observe that $\partial V_R(\phi,\chi)/\partial\phi=\partial V_R(\phi,\chi)/\partial\chi=0$, for $0<s<\infty$ and $0<\mu<\infty$, is achieved at $\phi_m=\chi_m=0$, resulting in the curvature of the real part of $V_R$ having the same size in the $\phi$ or $\chi$-direction, and the vacuum energy density is $V_R(0,0)=0$. 
%However, notice that for a more general plateau written as 
%\begin{eqnarray}
%    P_g(x)=\frac{1+\delta}{1+(x/\mu)^2},
%\end{eqnarray}
%the minimum $\partial V_R(\phi,\chi)/\partial\phi=\partial V_R(\phi,\chi)/\partial\chi=0$ remains at $\phi_m=\chi_m=0$. Nevertheless, the vacuum energy density now has the value $V(0,0)=\delta V_0$. 
%For simplicity, hereafter, one assumes the plateau function is given by equation \eqref{eq:plateau}. 
%In this case, one expects that at the end of inflation (at $t=t_{\rm end}$), $P(x)\rightarrow 1$. Then, considering $0<\mu<1$ and $\phi^2+\chi^2<1$, the asymptotic limit $P(x)\rightarrow 1$ can be achieved if $\phi^2+\chi^2<\mu^{2}$. Therefore, at the end of inflation
%\begin{eqnarray}
%    V_0\bigl[1-P(x)\bigr]\rightarrow 0.
%\end{eqnarray}

%The above result is achieved at the end of inflation, where $V_R(\phi,\chi)=0$ (at $t_{\rm end}$) $\phi\approx \chi\ll 1$. In this case, at $V(\phi,\chi)=0$, one has
%\begin{eqnarray}\label{eq:phimenor}
%    \chi(t_{\rm end})\approx \left[\frac{2}{\lambda}(m^2+\varepsilon_\phi+\varepsilon_\chi) \right]^{1/2}= \left[\frac{2}{\lambda}(m^2+2) \right]^{1/2},~~ \phi\ll\chi
%\end{eqnarray}
%while for the second case, one has
%\begin{eqnarray}\label{eq:phiigual}
%   \phi(t_{\rm end})\approx  \chi(t_{\rm end})\approx \frac{m}{\sqrt{\lambda}}. 
%\end{eqnarray}

The quartic self-coupling constant $\lambda$ is expected to be weak to ensure a sub-Planckian potential energy ($V_R<\!\!<M_P^4$), ensuring that quantum gravity effects can be neglected in the context \cite{baumann.2012}.  The mass parameter is also taken as a small $m=10^{11}-10^{13}$ GeV: $m\sim 10^{-8}-10^{-6}\,M_P$ \cite{stein.2021}. For simplicity in numerical calculations, hereafter, it is assumed $\lambda=10^{-13}\,M_P$, and $m=10^{-6}\,M_P$. %Taking this value into consideration, one obtains the following result for equation \eqref{eq:phiigual}
%\begin{eqnarray}
%   \chi^2(t_{\rm end})&\approx& 0.01 \rightarrow \phi^2(t_{\rm end})+\chi^2(t_{\rm end})\approx 0.02 < \frac{1}{\mu^2},
%\end{eqnarray}
%As can be quickly verified, both results corroborate the assumption $P(x)\rightarrow 1$ at $t_{\rm end}$. Nevertheless, this condition is more favored by the assumption $\phi(t_{\rm end})\ll\chi(t_{\rm end})$.

The parameter $V_0$ sets the overall energy scale of the real part of the inflaton potential, fixed by the observed amplitude of the primordial scalar power spectrum, as noted by the COBE/Planck Collaboration \cite{planck.collab.2015}. For a given choice of dimensionless couplings $(\epsilon_\phi,\epsilon_\chi,\lambda,\dots)$, a reference normalization $V_{\rm ref}$ produces a reference amplitude $A_s^{\rm ref}$. Since the slow-roll parameter $\epsilon_V$ is invariant under a global rescaling of the potential, the correct normalization to the observed value $A_s$ is obtained by rescaling the height of the potential according to the definition
\begin{eqnarray}
 \nonumber   V_0=V_{\rm ref}\frac{A_s}{A_s^{\rm ref}},
\end{eqnarray}
where $V_{\rm ref}$ and $A_s^{\rm ref}$ are reference values. To fix the mass scale, we impose that the predicted amplitude of the primordial scalar power spectrum agrees with the value inferred from the Planck data \cite{linde.2025,pozo.2024,cook.2015}. Specifically, the dimensionless scalar power amplitude is defined as \cite{baumann.2015}
\begin{eqnarray}
\nonumber    A_s\equiv \mathcal{P}(k_*)\simeq \frac{1}{24\pi^2M_P^4}\left(\frac{V_{\rm eff}}{\epsilon_V}\right)_{k=k_*},
\end{eqnarray} 
in which $V_{\rm eff}$ is the effective potential, and $\epsilon_V$ is its slope, given by equation \eqref{eq:ev}.

It is important to stress that for $\epsilon_V<\!\!<1$, the field rolls slowly, implying that the potential energy dominates, thus furnishing accelerated expansion. At the pivot scale $k_*$ (during the slow-roll phase), the scalar power spectrum amplitude is given by \cite{baumann.2015}
\begin{eqnarray}
\nonumber   A_s\propto \frac{V_{\rm eff}}{\epsilon_V},
\end{eqnarray}
and $\epsilon_V$ remains unchanged with a global rescaling of the potential. Moreover, at $k_* = 0.05\,\text{Mpc}^{-1}$, one has $A_s \approx 2.085 \times 10^{-9}$  \cite{cook.2015}. 

The imaginary part of $V(\Phi)$ is defined by the product
\begin{eqnarray}\label{eq:imagpot}
    V_I(\phi,\chi)=(\varepsilon_\phi-\varepsilon_\chi)\phi\chi=\Delta\varepsilon\,\phi\chi,
\end{eqnarray}
which is the simplest purely imaginary bilinear contribution in the fields $\phi\chi$. The imaginary potential \eqref{eq:imagpot} enables independent tuning of dissipation (through the parameter $\Delta\epsilon$) without perturbing the inflationary background, which is predominantly determined by the real couplings and the non‑minimal coupling $\zeta$ (as shall be seen). Furthermore, note the absence of $V_0$ in this sector: while $V_0$ fixes the overall energy scale of the real inflationary plateau—being constrained by the observed scalar amplitude $A_s$—the parameter $\Delta\varepsilon$ governs the magnitude of the non‑Hermitian effects. 

%---------------------
\subsection{Diagonalization}

Since the potential is complex for $\varepsilon_\phi\neq\varepsilon_\chi$, the diagonalization of the quadratic sector necessarily involves complex field redefinitions.  Notice that the quartic term $\frac{\lambda}{4}(\phi^2+\chi^2)^2$ is invariant under real orthogonal rotations $(\phi,\chi)\rightarrow(\phi',\chi')$. However, the mixed term $i\Delta\varepsilon\,\phi\chi$ is purely imaginary for $\varepsilon_\phi\neq \varepsilon_\chi$, and its removal requires the application of a complex linear transformation; specifically, a hyperbolic transformation that does not leave the quantity $\phi^2 + \chi^2$ invariant. Therefore, the potential \eqref{eq:pot2} cannot be both diagonal and quadratic/quartic at the same time. We choose to keep the quadratic/quartic sector, defining
\begin{eqnarray}
\nonumber    A=\frac{\varepsilon_\phi+\varepsilon_\chi}{2}, ~~B=\frac{\varepsilon_\phi-\varepsilon_\chi}{2},
\end{eqnarray}
and rewriting the quadratic sector of the potential as
\begin{eqnarray}\label{eq:matrix}
\nonumber    V_2(\phi,\chi)&=&\left(A-\frac{m^2}{2}\right)\phi^2+\left(-A-\frac{m^2}{2}\right)\chi^2+2iB\phi\chi=\\
&=& \begin{pmatrix}
\phi & \chi
\end{pmatrix}
\begin{pmatrix}
A-\frac{m^2}{2} & iB\\
iB & -A-\frac{m^2}{2}
\end{pmatrix}
\begin{pmatrix}
\phi \\
\chi\\
\end{pmatrix}.
\end{eqnarray}

The corresponding mass eigenvalues related to the central matrix on the right-hand side (rhs) of \eqref{eq:matrix} are given by
\begin{eqnarray}
\nonumber    \mu_{\pm}=-\frac{m^2}{2}\pm\sqrt{\varepsilon_\phi\varepsilon_\chi},
\end{eqnarray}
being complex for $\varepsilon_\phi\varepsilon_\chi<0$, consistent with the fact that $V(\phi,\chi)$ is complex. Therefore, there is a base $(\psi,\eta)$ where we can write
\begin{eqnarray}
 \nonumber   V_2=\mu_+\psi^2+\mu_-\eta^2.
\end{eqnarray}

Using $\theta$ complex, the linear transformation
\begin{eqnarray}\label{eq:matrix_transf}
\nonumber    \begin{pmatrix}
        \phi\\
        \chi
    \end{pmatrix}=\begin{pmatrix}
        \cos\theta & -\sin\theta\\
        \sin\theta & \cos\theta
    \end{pmatrix}\begin{pmatrix}
        \psi\\
        \eta
    \end{pmatrix}=R(\theta)\begin{pmatrix}
        \psi\\
        \eta
    \end{pmatrix},
\end{eqnarray}
allows one to remove the imaginary term in the potential. Observe that the condition 
\begin{eqnarray}
\nonumber    \tan(2\theta)=\frac{2iB}{\left(A-\frac{m^2}{2}\right)-\left(-A-\frac{m^2}{2}\right)}=\frac{iB}{A}, ~~A\neq 0,
\end{eqnarray}
can be used by assuming the parameterization $2\theta=i\zeta$, $\zeta\in\mathbb{R}$ for $|B/A|<1$, since
\begin{eqnarray}
\nonumber   \tan(i\zeta)=i\tanh\zeta\rightarrow \tanh\zeta=\frac{B}{A}=\frac{\varepsilon_\phi-\varepsilon_\chi}{\varepsilon_\phi+\varepsilon_\chi},
\end{eqnarray}
implying
\begin{eqnarray}
\nonumber    \cos\theta=\cosh(\zeta/2), \qquad \sin\theta=i\sinh(\zeta/2).
\end{eqnarray}

Therefore, the explicit linear transformation is given by
\begin{eqnarray}
\nonumber\begin{cases}
    \phi(\psi,\eta)=\cosh(\zeta/2)\psi-i\sinh(\zeta/2)\eta,\\
    \chi(\psi,\eta)=i\sinh(\zeta/2)\psi+\cosh(\zeta/2)\eta
\end{cases}
\end{eqnarray}
allowing us to rewrite the complex potential \eqref{eq:pot2} in terms of $\psi$ and $\eta$
\begin{eqnarray}
\nonumber    V(\psi,\eta)=\mu_+\psi^2+\mu_-\eta^2+\frac{\lambda}{4}\bigl[\phi(\psi,\eta)^2+\chi(\psi,\eta)^2 \bigr]^2.
\end{eqnarray}

The quadratic mass matrix in equation \eqref{eq:matrix} represents the local approximation around a chosen background point (e.g., $r \ll r_0$ on the plateau). Away from this neighborhood, the effective Hessian acquires a field-dependent structure that varies according to the radial amplitude $r$. In such regions, tachyonic instabilities and genuinely non-Hermitian regimes can coexist, leading to complex mass eigenvalues that reflect both vacuum instability and dissipative dynamics~\cite{gorini.2004}. Nevertheless, the method of complex field rotation (diagonalization) provides a consistent framework for analyzing the quadratic sector and identifying the nature of the instabilities across the full field space.

%With the radial form factor $P(r)$, the quadratic mass matrix in equation \eqref{eq:matrix} should be understood as the local quadratic approximation around $r<\!\!<r_0$ (or around the chosen background point). Outside this neighborhood, the effective Hessian acquires a nontrivial dependence on the background field configuration. More generally, once the full Hessian of the potential is allowed to vary with the radial field amplitude $r$, tachyonic and genuinely non-Hermitian regimes can, in principle, coexist in different regions of field space away from the origin \cite{gorini.2004}. Nonetheless, the method of complex rotation remains the appropriate framework for characterizing the non-Hermitian quadratic form and identifying the corresponding instabilities.

%===================
\section{The Lagrangian Formulation}\label{sec:lagrangian}

We introduce a non-minimal coupling between the scalar field and the spacetime curvature, parameterized as
\begin{eqnarray}
 \nonumber   \left(\frac{M_P^2}{2}-\zeta \Phi\Phi^{*} \right)R,
\end{eqnarray}
where $R$ is the Ricci scalar curvature, and $\zeta$ is the dimensionless non-minimal coupling constant that controls the strength of the field-dependent modification to the effective gravitational coupling. For a conformally invariant massless scalar field, $\zeta=1/6$ \cite{wald.1984} (for details, see Appendix D of Ref. \cite{wald.1984}). In particular, if $\zeta\ne 1/6$, then particle production by the gravitational field is possible \cite{ford.1987}. In general, one expects $\zeta>0$, which means $\zeta$ screens the Planck mass at large field values, turning gravity stronger and enhancing Hubble friction, which tends to prolong slow roll. Conversely, $\zeta<0$ anti‑screens the Planck mass: The effective gravitational interaction becomes weaker at large field amplitudes, which in turn reduces the associated frictional damping and thereby makes it more difficult to sustain a prolonged period encompassing many e‑folds of expansion. %Moreover, the effective gravitational field... 

To ensure the positivity of the metric, the absence of ghosts implies the positivity of the effective gravitational coupling term
\begin{equation}\label{eq:ftheta}
F(\phi,\chi)=M_P^2-\zeta(\phi^2+\chi^2)>0,
\end{equation}
moreover, observe that the effective gravitational constant \cite{salopek.1989}
\begin{eqnarray}
\nonumber   G_{\rm eff}=F(\phi,\chi)^{-1}=\frac{1}{M_P^2-\zeta(\phi^2+\chi^2)},
\end{eqnarray}
decreases when the field amplitude grows for $\zeta<0$ \cite{tsujikawa.2000}. 
%\begin{eqnarray}
%    M_P^2-\zeta\bigl(\phi^2+\chi^2\bigr)>0.
%\end{eqnarray}

%\begin{equation}
%\Phi^{*}\Phi=\frac{1}{2}(\phi^2+\chi^2),\qquad
%\frac{M_P^2}{2}-\zeta\,\Phi^{*}\Phi
%=\frac{1}{2}\Bigl(M_P^2-\zeta(\phi^2+\chi^2)\Bigr)
%\equiv \frac{1}{2}F(\phi,\chi).
%\end{equation}

Using the complex potential given by equation \eqref{eq:pot2}, the CIF $\Phi$ can be coupled with gravity through the action
\begin{widetext}
\begin{eqnarray}\label{eq:action}
S=\int d^4x \sqrt{-g}\left[\frac{1}{2}\Bigl(M_p^2-2\zeta\,\Phi^{*}\Phi\Bigr)R
- g^{\mu\nu}\partial_\mu{\Phi}^{*}\partial_\nu\Phi
- V(\Phi) \right],
\end{eqnarray}
\end{widetext}
where $\sqrt{-g}=a(t)^3=a^3$, and $a(t)$ is the scale factor in the flat Friedmann-Robertson-Walker (FRW) background 
\begin{equation}
\nonumber ds^2=-dt^2+a(t)^2\delta_{ij}dx^idx^j,\,
H\equiv \frac{\dot a}{a},\,
R=6(2H^2+\dot H),
\end{equation}
with $H$ as the Hubble constant. Moreover, notice that
\begin{eqnarray}
\nonumber g^{\mu\nu}\partial_\mu \Phi^{*}\partial_\nu \Phi=
\frac12 g^{\mu\nu}\left(\partial_\mu\phi\,\partial_\nu\phi+\partial_\mu\chi\,\partial_\nu\chi\right).
\end{eqnarray}

It is important to emphasize that, throughout this work, spacetime geometry is determined solely by the real part of the energy–momentum tensor, while the imaginary part acts as a signature of non-conservative dynamics.

%============
\subsection{Tachyons}

A tachyon ($m^2 < 0$) indicates an instability of the vacuum configuration around which the field equations are expanded~\cite{nielsen.1978,gorini.2004}. In the context of inflation, tachyonic modes are not pathological; they indicate that the chosen background is unstable, and the field will roll toward the true minimum~\cite{sen.2002}. This mechanism has been successfully implemented in models of hybrid inflation, as well as in other theoretical frameworks in which vacuum instability constitutes the primary driver of the dynamical evolution \cite{padmanabhan.2002}. In this work, tachyonic eigenmodes can appear in certain regions of field space alongside genuinely non-Hermitian (complex-mass) regimes dominated by dissipation, thereby providing a framework for modeling reheating.

In the present analysis, tachyonic eigenmodes arise under the assumption that \(M\) denotes the conventional mass parameter appearing in the quadratic contribution to the potential, which is generically expressed as $V_2 = (M^2/2)\sigma^2$, where $\sigma$ is some real scalar field. Comparing this term with the preceding results, one obtains
\begin{eqnarray}
\nonumber    M_\pm^2=2\mu_\pm=-m^2 \pm2\sqrt{\varepsilon_\phi\varepsilon_\chi}.
\end{eqnarray}

Considering $\varepsilon_\phi\varepsilon_\chi>0$,  the presence of tachyons implies $M_\pm^2<0$, leading to the following conditions:
\begin{eqnarray}
\label{eq:inel1}   M_+^2&=&-m^2+2\sqrt{\varepsilon_\phi\varepsilon_\chi}<0,\\
\label{eq:inel2}    M_-^2 &=&-m^2-2\sqrt{\varepsilon_\phi\varepsilon_\chi}<0.
\end{eqnarray}

From the inequality \eqref{eq:inel1}, the condition $M_+^2<0$  requires $m^2>2\sqrt{\varepsilon_\phi\varepsilon_\chi}$, whereas from the inequality \eqref{eq:inel2}, it is always true for $m^2>0$. Therefore, if $m^2>0$ and $\varepsilon_\phi\varepsilon_\chi>0$, the inequality \eqref{eq:inel2} always leads to a tachyonic eigenmode around the origin of the potential. However, considering $\varepsilon_\phi\varepsilon_\chi<0$, $M_\pm^2$ is complex, resulting in a non-Hermitian system (a dissipative system). 
%These two regimes ($\varepsilon_\phi\varepsilon_\chi>0$ and $\varepsilon_\phi\varepsilon_\chi<0$) are mutually exclusive: tachyonic instabilities arise only for real mass eigenvalues, while complex masses correspond to genuinely non-Hermitian, dissipative dynamics. 
Furthermore, notice that when $M^2$ is complex, it can be used to represent an unstable particle \cite{S.Willenbrock.2024} (see also Ref. \cite{sergeenko.2014} and references therein)
\begin{eqnarray}
\nonumber    M^2=M_R^2+iM_R\Gamma,
\end{eqnarray}
where $M_R$ is the central energy (the measurable mass) of the unstable particle, and $\Gamma$ is the decay width of the particle: for $\Gamma = 0$, the particle is stable, whereas for $\Gamma > 0$ the particle becomes unstable. For unstable particles, quantum corrections displace the propagator pole from the real energy axis into the complex plane, characterized by a central energy $M_R$ (the particle mass) and a decay width $\Gamma$~\cite{S.Willenbrock.2024}. In the present model, complex mass eigenvalues play an analogous role, encoding both the energy scale of the mode and its dissipative lifetime in the expanding Universe, thereby providing a geometric mechanism for effective particle decay without explicit couplings to daughter fields. 

%Considering unstable particles, one must acknowledge that interactions and loop corrections can cause the propagator pole to be displaced from the real energy axis into the complex energy plane. The location of this complex pole is characterized by the parameters \(M_{R}\) and \(\Gamma\) \cite{S.Willenbrock.2024}.

%===============
\subsection{Ostrogradsky Ghosts}

Ostrogradsky ghosts correspond to degrees of freedom associated with an ‘incorrect sign’ in the kinetic term of the Lagrangian. Actually, the Ostrogradsky theorem states that a non-degenerate Lagrangian with finite higher-order time derivatives results in a Hamiltonian that is unbounded from below \cite{klein.roest.2016,ostrogradsky.1850}.  Phantom or ghost scalar models \cite{ludwick.2017,vikman.2005} belong to a class of quintessence models where $w_R<-1$. In general, ghost scalar models present instabilities and violate energy conditions \cite{singh.2003}.

The kinetic sector can be kept canonically normalized by construction in the transformed variables. %since $R(\theta)^T R(\theta)=I$, where $R^T(\theta)$ is the transpose of $R(\theta)$ and $I$ is the identity in $M_{2\times2}$. 
Then, from the action \eqref{eq:action}, the Lagrangian representing the kinetic term is
\begin{eqnarray}
\nonumber\mathcal{L}_K&=&\frac{1}{2}\partial_\nu\phi\partial^\nu\phi + \frac{1}{2}\partial_\nu\chi\partial^\nu\chi=\\
\nonumber   &=&\frac{1}{2}\partial_\nu\psi\partial^\nu\psi + \frac{1}{2}\partial_\nu\eta\partial^\nu\eta,
\end{eqnarray}
implying the non-emergence of spurious negative signs in the model: there are no ghosts. Therefore, the presence of a negative square mass corresponds to tachyonic instabilities, and the eventual $\varepsilon_\phi\varepsilon_\chi<0$ is due to the dissipative nature of the system.

%=================
\subsection{Equations of Motion}

The equations of motion for $\phi$ and $\chi$ can be written in a compact form as
\begin{eqnarray}\label{eq:eom_compact}
\nabla_\mu\nabla^\mu \pi_l-\frac{\partial V(\phi,\chi)}{\partial \pi_l}+\zeta\,R\,\pi_l=0,
\qquad l=1,2,
\end{eqnarray}
where $\pi_1=\phi$ and $\pi_2=\chi$. The term $\partial V(\phi,\chi)/\partial \pi_l$ can be written explicitly as
%\begin{eqnarray}
%    \frac{\partial V_R(\phi,\chi)}{\partial\pi_j}=P(r)\frac{\partial V_{p}}{\partial\pi_j}+\bigl(V_{p}-V_f\bigr)\frac{d P(r)}{dr}\frac{\partial r}{\partial\pi_j},~~ \frac{\partial r}{\partial\pi_j}=\frac{\pi_j}{r}
%\end{eqnarray}
\begin{eqnarray}
\nonumber\frac{\partial V}{\partial\phi} &=&
V_0\left\{P(r)\,\frac{\partial V_{p}}{\partial\phi}
+\bigl[V_{p}(\phi,\chi)-1]\,P'(r)\,\frac{\phi}{r}\right\}+\\
&+&i\Delta\varepsilon\chi, \label{eq:dVphi_mod}\\
\nonumber\frac{\partial V}{\partial\chi} &=&
V_0\left\{P(r)\,\frac{\partial V_{p}}{\partial\chi}
+\bigl[V_{p}(\phi,\chi)-1]\,P'(r)\,\frac{\chi}{r}\right\}+\\
&+&i\Delta\varepsilon\phi,\label{eq:dVchi_mod}
\end{eqnarray}
with
\begin{eqnarray}
\nonumber\frac{\partial V_{ p}}{\partial\phi}
&=&\bigl[(\varepsilon_\phi+\varepsilon_\chi-m^2)
+\lambda(\phi^2+\chi^2)\bigr]\phi,\\
\nonumber \frac{\partial V_{ p}}{\partial\chi}
&=&\bigl[-(\varepsilon_\phi+\varepsilon_\chi+m^2)
+\lambda(\phi^2+\chi^2)\bigr]\chi.
\end{eqnarray}

Therefore, the corresponding equation of motion \eqref{eq:eom_compact} for $\pi_j$ is the usual Klein-Gordon equation in curved spacetime with a non-minimal coupling term. 
%and write the non-minimal coupling function as
%\begin{equation}
%F(\phi,\chi)=M_P^2-\zeta(\phi^2+\chi^2).
%\end{equation}
%The complex potential is decomposed as $V(\phi,\chi)=V_R(\phi,\chi)+iV_I(\phi,\chi)$ with
%\begin{equation}
%V_I(\phi,\chi)=(\varepsilon_\phi-%\varepsilon_\chi)\phi\chi,
%\qquad
%r\equiv\sqrt{\phi^2+\chi^2},
%\end{equation}
%and the real part is taken to be the EFT-motivated form-factor deformation
%\begin{equation}
%V_R(\phi,\chi)=V_{ p}(\phi,\chi)\,P(r)+V_f\,[1-P(r)].
%\end{equation}
%The field derivatives entering the equations of motion are
%\begin{eqnarray}
%\frac{\partial V}{\partial\phi} &=&
%P(r)\,\frac{\partial V_{ p}}{\partial\phi}
%+\bigl[V_{ p}(\phi,\chi)-V_f\bigr]\,P'(r)\,\frac{\phi}{r}
%+i(\varepsilon_\phi-\varepsilon_\chi)\chi, \label{eq:dVphi_FRW}\\
%\frac{\partial V}{\partial\chi} &=&
%P(r)\,\frac{\partial V_{ p}}{\partial\chi}
%+\bigl[V_{ p}(\phi,\chi)-V_f\bigr]\,P'(r)\,\frac{\chi}{r}
%+i(\varepsilon_\phi-\varepsilon_\chi)\phi. \label{eq:dVchi_FRW}
%\end{eqnarray}
%\begin{equation}
%\frac{\partial V_{ p}}{\partial\phi}
%=(\varepsilon_\phi+\varepsilon_\chi-m^2)\phi+\lambda\phi(\phi^2+\chi^2),
%\qquad
%\frac{\partial V_{ p}}{\partial\chi}
%=-(\varepsilon_\phi+\varepsilon_\chi+m^2)\chi+\lambda\chi(\phi^2+\chi^2).
%\end{equation}
The homogeneous field equations in the Jordan frame take the form
\begin{eqnarray}\label{eq:eom_FRW}
\begin{cases}
\ddot\phi+3H\dot\phi-\zeta R\,\phi+\frac{\partial V}{\partial\phi}=0,\\
\ddot\chi+3H\dot\chi-\zeta R\,\chi+\frac{\partial V}{\partial\chi}=0,
\end{cases}
\end{eqnarray}
where $\partial V/\partial\phi$ and $\partial V/\partial\chi$ are given in
Eqs. \eqref{eq:dVphi_mod} and \eqref{eq:dVchi_mod}. It is often convenient to rewrite equations \eqref{eq:eom_FRW} in terms of the number of e-folds $N\equiv\ln a$ (now prime denotes $d/dN$):
\begin{eqnarray}\label{eq:eom_N}
\begin{cases}
\phi''+(3-\epsilon_H)\phi'-\frac{\zeta R}{H^2}\phi+\frac{1}{H^2}\frac{\partial V}{\partial\phi}=0,\\
\chi''+(3-\epsilon_H)\chi'-\frac{\zeta R}{H^2}\chi+\frac{1}{H^2}\frac{\partial V}{\partial\chi}=0,
\end{cases}
\end{eqnarray}
with $\epsilon_H\equiv-\dot H/H^2$ and $R/H^2=6(2-\epsilon_H)$. Furthermore, $N_{\rm end}\in[50,60]$ corresponds to modes exiting the Hubble radius $\sim 50-60$ e-folds before the end of inflation.  
%Assuming a flat FRW universe with signature $(-,+,+,+)$, one has the following system of equations
%\begin{eqnarray}
%\begin{cases}\ddot{\phi}+3H\dot{\phi}
%+\nonumber\frac{\partial V(\phi,\chi)}{\partial\phi}
%+\zeta\,R\,\phi=0,\\
%\ddot{\chi}+3H\dot{\chi}
%+\frac{\partial V(\phi,\chi)}{\partial\chi}
%+\zeta\,R\,\chi=0.
%\end{cases}
%\end{eqnarray}
%and, explicitly, one writes
%\begin{eqnarray}\label{eq:eom_frw_explicit}
%\begin{cases}\ddot{\phi}+3H\dot{\phi}
%+\Big[(\varepsilon_\phi+\varepsilon_\chi-m^2)+\lambda(\phi^2+\chi^2)+\zeta R\Big]\phi
%+i\Delta\varepsilon\chi=0,\\
%\ddot{\chi}+3H\dot{\chi}
%+\Big[-(\varepsilon_\phi+\varepsilon_\chi+m^2)+\lambda(\phi^2+\chi^2)+\zeta R\Big]\chi
%+i\Delta\varepsilon\phi=0.
%\end{cases}
%\end{eqnarray}
%where $\sqrt{-g}=a(t)^3=a^3$, $H=\dot{a}/a$ stands for the Hubble parameter, and $R=6(2H^2+\dot{H})$. 
As usual, the energy-momentum tensor is written in a spatially flat FRW spacetime as
\begin{eqnarray}
    \nonumber T^{\alpha\beta}=\frac{\partial\mathcal{L}}{\partial(\partial_\alpha\pi_j)}\partial^{\alpha}\pi_j-g^{\alpha\beta}\mathcal{L},
\end{eqnarray}
and, consequently, the effective energy density and pressure associated with the homogeneous scalar sector in the Jordan frame are given by
\begin{eqnarray}
\nonumber \rho = \rho(\phi,\chi)&=& \frac{1}{2}\left(\dot\phi^{\,2}+\dot\chi^{\,2}\right)
+V(\phi,\chi)\\
&-&3H\dot F(\phi,\chi), \label{eq:rho_JF}\\[2mm]
\nonumber p =p(\phi,\chi)&=& \frac{1}{2}\left(\dot\phi^{\,2}+\dot\chi^{\,2}\right)
-V(\phi,\chi)+\\
&+&\ddot F(\phi,\chi)+
+ 2H\dot F(\phi,\chi), \label{eq:press_JF}
\end{eqnarray}
in which
\begin{eqnarray}
%F(\phi,\chi)=M_P^2-\zeta(\phi^2+\chi^2),
%\qquad
\nonumber \dot F=-2\zeta(\phi\dot\phi+\chi\dot\chi),\,
\ddot F=-2\zeta(\dot\phi^{\,2}+\dot\chi^{\,2}+\phi\ddot\phi+\chi\ddot\chi).
%\label{eq:F_dots}
\end{eqnarray}

Given that the potential energy $V(\phi,\chi)\in\mathbb{C}$, the corresponding energy density and pressure also belong to $\mathbb{C}$ as well. Then, one writes
\begin{eqnarray}
\nonumber\rho=\rho_R+i\rho_I,~~
\nonumber p=p_R+ip_I,
\end{eqnarray}
with
\begin{eqnarray}
\nonumber \rho_R &=& \frac{1}{2}\left(\dot\phi^{\,2}+\dot\chi^{\,2}\right)+V_R-3H\dot F,\,\rho_I = V_I,\\
\nonumber p_R &=& \frac{1}{2}\left(\dot\phi^{\,2}+\dot\chi^{\,2}\right)-V_R+\ddot F+2H\dot F,\,p_I = -V_I.
\end{eqnarray}

The real part of the complex pressure, $p_R$, represents the physical pressure that sources the Friedmann acceleration. In contrast, the imaginary part, $p_I$, is not an observable quantity in standard cosmology; rather, it is an effective parameter quantifying the non-conservative nature of the system. Specifically, $p_I$ (and similarly, the imaginary component of the energy density) arises from the field-dependent coupling to gravity and encodes the rate of energy transfer away from the observable inflaton sector. The real part of the energy density, $\rho_R$, determines the expansion of the Universe through the Friedmann equations. 

It is important to stress that the complex structure of $\rho$ does not represent a new form of matter. It is a mathematical tool for describing open-system dynamics in which the inflaton sector exchanges energy with unobserved degrees of freedom (the reheating bath) as it approaches the end of inflation.
%In the Jordan frame, this energy exchange appears as a departure from the standard conservation law $\dot{\rho} + 3H(\rho + p) = 0$, but it is fully accounted for in the Einstein frame through minimal coupling and the conservation of the total energy-momentum tensor. 

The modified Friedmann equations take the usual scalar-tensor form,
\begin{eqnarray}
\begin{cases}\label{eq:friedmann}
3F H^2 &= \rho,\\
-2F\dot H &= \rho+p-\ddot F + H\dot F,
\end{cases}
\end{eqnarray}
and, equivalently, combining Eqs.~(\ref{eq:rho_JF})--(\ref{eq:press_JF}) yields
\begin{equation}
\nonumber-2F\dot H=\dot\phi^{\,2}+\dot\chi^{\,2}-H\dot F.
\label{eq:Hdoteq}
\end{equation}

%resulting in the effective energy density and pressure given, respectively, by 
% \begin{eqnarray}
%\label{eq:rho}\rho &=&
%\frac12(\dot\phi^2+\dot\chi^2) +6\zeta H(\phi\dot\phi+\chi\dot\chi)
%+3\zeta H^2(\phi^2+\chi^2)+V(\phi,\chi),\\
%\label{eq:press}p&=&
%\frac12(\dot\phi^2+\dot\chi^2)-2\zeta(\phi\ddot\phi+\chi\ddot\chi)
%-4\zeta H(\phi\dot\phi+\chi\dot\chi) -\zeta(2\dot H+3H^2)(\phi^2+\chi^2)-V(\phi,\chi).
%\end{eqnarray}

No higher-order time derivatives appear in the equations of motion; the acceleration terms in the pressure originate from the variation of the non-minimal coupling and do not signal Ostrogradsky instabilities. From the system of equations \eqref{eq:eom_FRW} (or \eqref{eq:eom_N}), one finds that the energy balance equation can be expressed as
\begin{equation}
\nonumber\dot\rho+3H(\rho+p)=\frac{1}{2}\dot F\,R,
\label{eq:continuity_source}
\end{equation}
which makes explicit the energy exchange between the scalar sector and the spacetime curvature induced by the non-minimal coupling \cite{lobo.2025}.
Using the effective gravitational coupling term given by equation \eqref{eq:ftheta},
%\begin{equation}\label{eq:ftheta}
%F(\phi,\chi)=M_P^2-\zeta(\phi^2+\chi^2),
%\end{equation}
one obtains the modified Friedmann equations in a spatially flat FRW background, written as
\begin{eqnarray}\label{eq:friedmann1}
\begin{cases}
3F H^2 =
\frac{1}{2}\left(\dot{\phi}^2+\dot{\chi}^2\right)
+ V(\phi,\chi)
- 3H\dot{F},\\
-2F\dot{H} = \dot{\phi}^2+\dot{\chi}^2
+ \ddot{F} - H\dot{F}.
\end{cases}
\end{eqnarray}
%where
%\begin{equation}
%\nonumber \dot{F}=-2\zeta(\phi\dot{\phi}+\chi\dot{\chi}), \qquad
%\ddot{F}=-2\zeta(\dot{\phi}^2+\dot{\chi}^2+\phi\ddot{\phi}+\chi\ddot{\chi}).
%\end{equation}

The system of equations \eqref{eq:friedmann1}, together with the scalar equations of motion \eqref{eq:eom_FRW}, forms a closed system governing the cosmological evolution in the Jordan frame. In the Einstein frame, $\tilde g_{\mu\nu}=\Omega^2 g_{\mu\nu}$, $\Omega^2=F(\phi,\chi)/M_P^2>0$ (ghost-free), the scalar fields are minimally coupled, and the standard conservation law 
\begin{eqnarray}
\nonumber    \dot{\tilde\rho}+3\tilde H(\tilde\rho+\tilde p)=0
\end{eqnarray}
holds. This shows that the model is consistent with scalar–tensor gravity rather than with \textit{ad hoc} dissipation.

It is important to emphasize that Eqs. \eqref{eq:rho_JF}–\eqref{eq:press_JF} provide a covariant and gravitationally self-consistent description of the CIF decay that does not require the introduction of additional fields. In fact, this framework is closely analogous to the standard treatment of unstable particles in QFT \cite{S.Willenbrock.2024}, while being explicitly embedded in a cosmological context.

%Furthermore, the term on the right-hand side can be interpreted as an effective source, 
%\[
%Q = 6\zeta H\bigl(\dot{\phi}^2 + \dot{\chi}^2\bigr).
%\]

%In the Einstein frame, defined by the conformal transformation $\tilde g_{\mu\nu}=\Omega^2 g_{\mu\nu}$ with $\Omega^2=1-\zeta(\phi^2+\chi^2)/M_p^2$, the scalar fields are minimally coupled to gravity. Denoting by $\tilde\rho$ and $\tilde p$ the corresponding Einstein-frame energy density and pressure of the canonically normalized fields, one recovers the standard conservation law
%\begin{equation}
%\frac{d\tilde\rho}{d\tilde t}
%+3\tilde H(\tilde\rho+\tilde p)=0,
%\end{equation}
%so that energy conservation is restored once the gravitational sector is fully disentangled from the scalar dynamics. Thus, non-conservation is frame-dependent: in the Jordan frame, energy exchanges with gravity, while in the Einstein frame, the standard conservation law holds. This shows that the model aligns with scalar–tensor gravity rather than \textit{ad hoc} dissipation.

%============================
\section{Equation of State and Observational Data}\label{sec:eos}

%##############
\subsection{Equation of state}

To characterize the non-conservative dynamics, we define a complex, time-dependent equation of state as
\begin{equation}
w=w(t) = \frac{p(t)}{\rho(t)} = w_R + i w_I,
\label{eq:w_def}
\end{equation}
where $w_R$ and $w_I$ are the real and imaginary parts, respectively. Using the real and imaginary parts of $p$ and $\rho$, it is possible to rewrite $w_R$ and $w_I$ as
\begin{eqnarray}
\nonumber w_R &=& \frac{p_R\rho_R+p_I\rho_I}{\rho_R^2+\rho_I^2},
\label{eq:wR}\\[2mm]
\nonumber w_I &=& \frac{p_I\rho_R-p_R\rho_I}{\rho_R^2+\rho_I^2}.
\label{eq:wI}
\end{eqnarray}

The real part $w_R$ governs the effective gravitational interaction and determines the background expansion. As the inflaton evolves under the complex potential, the system passes through distinct dynamical (non-static) regimes characterized by $w_R$. In general, during the inflationary phase ($w_R \approx -1$), the system shows quasi-de Sitter expansion, closely resembling $\Lambda$CDM, with an almost constant negative real pressure that sustains accelerated expansion. In the phantom-like phase ($w_R < -1$), a transient over-acceleration arises when the imaginary components become significant, signaling rapid energy transfer from the inflaton to the reheating bath. This brief regime occurs near the reheating transition. Finally, in the reheating phase ($w_R > -1/3$), the EoS becomes less negative, indicating the onset of radiation domination as energy dissipation prevails and the inflaton decays into Standard Model particles.

In contrast, the imaginary part, $w_I$, directly measures the non-Hermitian nature of the inflaton sector and naturally relates to decay and reheating. Within this framework, a nonzero value of \(w_I\) indicates that the CIF does not behave as a closed fluid but instead exchanges energy with unobserved degrees of freedom. This transfer of energy constitutes the mechanism responsible for initiating the reheating phase.

%================
\subsection{Comparison to Observational Data}\label{sec:comp}

The viable parameter regions identified satisfy $N_{\rm end}\in[50,60]$ \cite{liddle.2003,german.2023} and ensure a standard inflationary epoch: the system exhibits a quasi-de Sitter phase ($w_R \simeq -1$) with non-Hermitian effects strongly suppressed ($|w_I| \ll 1$) at horizon exit. 

For a direct comparison with CMB observations, it is convenient to map the two-field dynamics in the Einstein frame to an effective single canonical inflaton $\varphi_{\rm eff}$. The conformal rescaling %\tilde g_{\mu\nu} = \Omega^2 g_{\mu\nu}$, with $\Omega^2 = F(\phi,\chi)/M_P^2$, 
renders the gravitational sector minimally coupled, and the background trajectory defines a radial effective field $\varphi_{\rm eff}(N)$ along which the real part of the potential generates an effective potential $V_{\rm eff}(\varphi_{\rm eff})$. In this one–field description, the standard potential slow-roll parameters are
\begin{equation}
\nonumber \epsilon_V \equiv \frac{M_P^2}{2}\left(\frac{V_{\rm eff}'}{V_{\rm eff}}\right)^2,
\qquad
\nonumber \eta_V \equiv M_P^2\,\frac{V_{\rm eff}''}{V_{\rm eff}},
\end{equation}
evaluated at a pivot e-fold $N_*\equiv N_{\rm end}\in[50,60]$. To leading order in slow roll, the spectral tilt, tensor-to-scalar ratio, and scalar amplitude are then given by
\begin{eqnarray}
\nonumber n_s \simeq 1 - 6\epsilon_{V*} + 2\eta_{V*},\,
r \simeq 16\,\epsilon_{V*},\,
A_s \simeq \frac{V_{\rm eff*}}{24\pi^2 M_P^4\,\epsilon_{V*}},
\end{eqnarray}
where the subscript $*$ denotes evaluation at $N_*$.
\begin{table}[t]
\centering
\setlength{\tabcolsep}{4pt}
\renewcommand{\arraystretch}{1.1}
\small
\begin{tabular}{lccc}
\hline\hline
 & $\zeta=0.1626$ & $\zeta=0.0001$ & $\zeta=-0.0053$ \\
 & $(\mu=0.1100)$ & $(\mu=0.1011)$ & $(\mu=0.1000)$ \\
\hline
$N_{\rm end}$      & 55.46           & 51.41           & 52.10 \\
$\epsilon_V$       & $4.120\times10^{-5}$ & $2.672\times10^{-5}$ & $2.612\times10^{-5}$ \\
$|\eta_V|$         & $1.952\times10^{-2}$ & $1.548\times10^{-2}$ & $1.537\times10^{-2}$ \\
$n_s$              & 0.961           & 0.969           & 0.969 \\
$r$                & 0.0007          & 0.0004          & 0.0004 \\
$V_{\rm eff}$      & $2.023\times10^{-11}$ & $1.313\times10^{-11}$ & $1.284\times10^{-11}$ \\
$A_s^{\rm ref}$    & 103.0           & 158.8           & 162.4 \\
$\chi^2$           & 1.146           & 0.936           & 1.006 \\
\hline\hline
\end{tabular}
\caption{Best-fit parameters of the model for three representative choices of the nonminimal coupling $\zeta$ and radial scale $\mu$. In all cases, the asymmetry parameter is fixed to $\Delta\varepsilon = 1.0$.}
\label{tab:comparison}
\end{table}

%\begin{table}[t!]
%    \centering
%    \begin{tabular}{c|c|c|c}
%Parameter      & $\zeta=0.1626$ ($\mu=0.1100$) & $\zeta=0.0001$ ($\mu=0.1011$) & $\zeta = -0.0053$ ($\mu=0.1000$) \\
%\hline
%$N_{\rm end}$  & 55.46                & 51.41 & 52.10     \\
%$\epsilon_V$   & 4.120$\times 10^{-5}$ & 2.672$\times10^{-5}$ & 2.612$\times 10^{-5}$ \\
%$|\eta_V|$     & 1.952$\times10^{-2}$ & 1.548$\times10^{-2}$ & 1.537$\times 10^{-2}$ \\ 
%$n_s$          & 0.961                &969 & 0.969  \\
%$r$            & 0.0007               &0.0004 & 0.0004   \\
%$V_{\rm eff}$  & 2.023$\times10^{-11}$&1.313$\times10^{-11}$ & 1.284$\times 10^{-11}$ \\
%$A_s^{\rm ref}$& 103.0                &158.8 & 162.4    \\
%$\chi^2$       & 1.146                &0.936 & 1.006    \\
%\hline
%     \end{tabular}
%    \caption{Parameters obtained from the model adopted. In all cases, $\Delta\varepsilon=1.0$.}
%    \label{tab:comparison}
%\end{table}

Using these expressions, the comparison with CMB data is quantified by the goodness-of-fit function
\begin{eqnarray}
\nonumber\chi^2_{\rm CMB} &\equiv&
\left(\frac{n_s-n_s^{\rm obs}}{\sigma_{n_s}}\right)^2+
\left(\frac{\ln A_s-\ln A_s^{\rm obs}}{\sigma_{\ln A_s}}\right)^2
+\\ \nonumber &+& \Theta(r-r_{\rm max})
\left(\frac{r-r_{\rm max}}{\sigma_r}\right)^2,
\label{eq:chi2cmb_eff}
\end{eqnarray}
where $n_s^{\rm obs}$, $A_s^{\rm obs}$, and $r_{\rm max}$ denote the observational central values and the current upper bound on primordial tensors, while $\sigma_{n_s}$, $\sigma_{\ln A_s}$, and $\sigma_r$ are the corresponding uncertainties. In practice, the overall height of $V_{\rm eff}$ is fixed once at a reference point by matching $A_s$ to the observed scalar amplitude, and this normalization is kept fixed when scanning over the parameters $\zeta$ and $\Delta\varepsilon$.

Table \ref{tab:comparison} exhibits the parameters obtained within the present model. The best $\chi^2$-parameter values were obtained for the following pairs: $(\zeta,\mu)=(0.1626,0.1100)$, $(\zeta,\mu)=(0.0001,0.1011)$, and $(\zeta,\mu)=(-0.0053,0.1000)$. The resulting $n_s$ are, respectively, $n_s\approx 0.961$, $n_s\approx 0.969$, and $n_s\approx 0.969$, which are in good agreement with the observed $n_s=0.965\pm0.004$ value from the Planck Collaboration \cite{planck.collab.2015}. For each $n_s$, one has the number of e-folds $N_{\rm end}=55.46$, $N_{\rm end}=51.41$, and $N_{\rm end}=52.10$ that lie within the expected range $N_{\rm end}\approx 50\sim 60$ \cite{liddle.2003,german.2023}.

\begin{figure}
    \centering
    \includegraphics[width=0.5\linewidth]{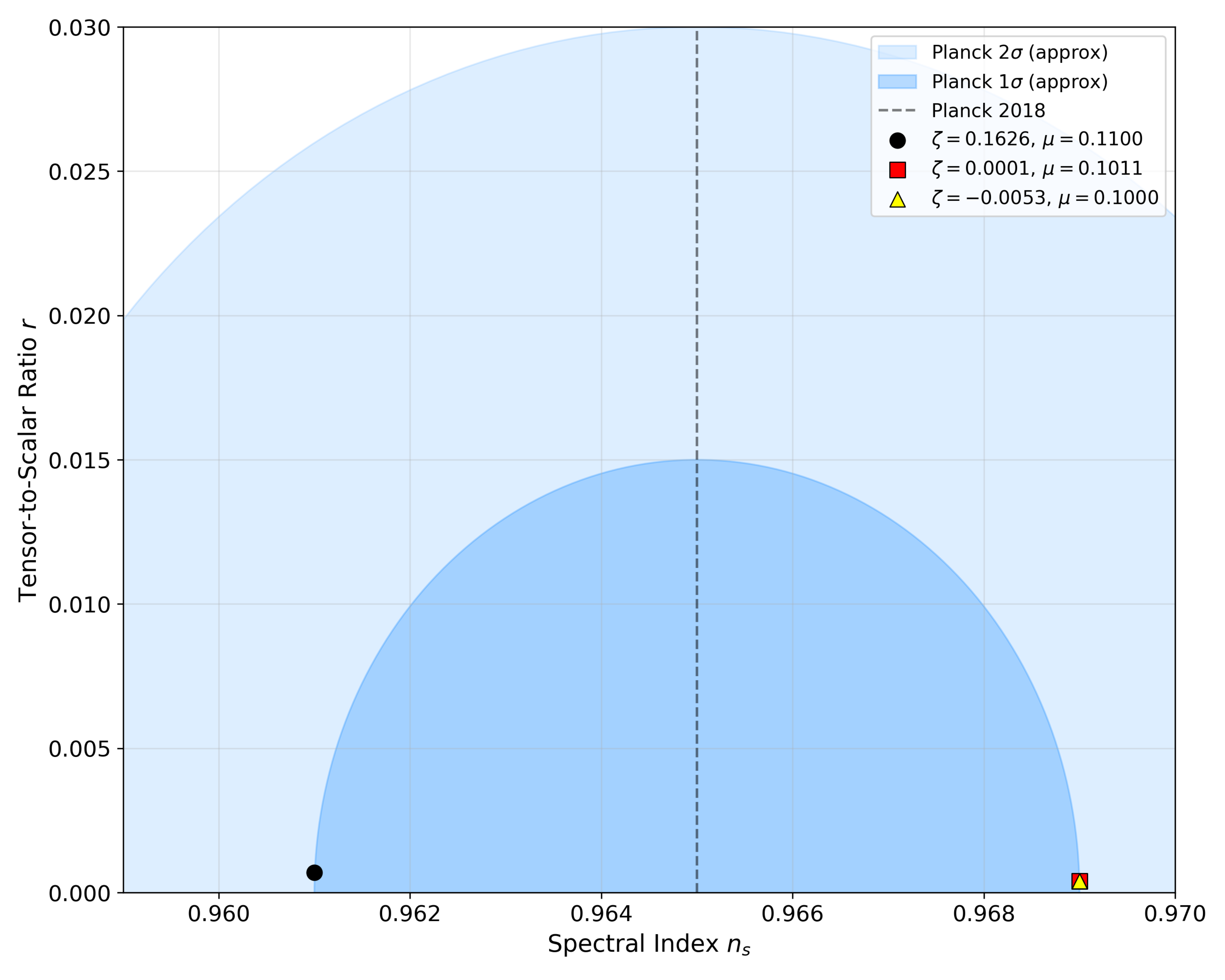}
\caption{Spectral index $n_s$ versus tensor-to-scalar ratio $r$ for the three best-fit parameter sets listed in Table \ref{tab:comparison} The points lie in the region $n_s \simeq 0.961-0.969$ and $r<0.06$, well inside the 1$\sigma$ and 2$\sigma$ Planck 2018 confidence contours (blue regions), showing that the complex inflaton model with nonminimal coupling is consistent with current CMB constraints.}
    \label{fig:bananaplot}
\end{figure}

As can be seen from Table \ref{tab:comparison}, the tensor-to-scalar ratio is far below $r=0.06$, as expected by the Planck Collaboration, as shown in Figure \ref{fig:bananaplot}, where the spectral index is plotted versus the tensor-to-scalar ratio in the ($n_s-r_s$)-plane. The points fall in the region $n_s\approx 0.961-0.969$ and $r<0.06$, implying the model predicts a spectrum consistent with Planck data, largely falling within 1$\sigma$ and 2$\sigma$ (the blue contours). Moreover, the model predicts an extremely low amplitude of primordial gravitational waves (tensor perturbations), yielding a tensor-to-scalar ratio that lies significantly below the current observational upper bound ($r < 0.06$). 

%%%%%%%%%%%%%%%%%
\section{The non-Hermitian sector}

%%%%%%%%%%%%%
\subsection{The number of e-folds}

\begin{figure}[t!]
    \centering
    \includegraphics[width=0.49\linewidth]{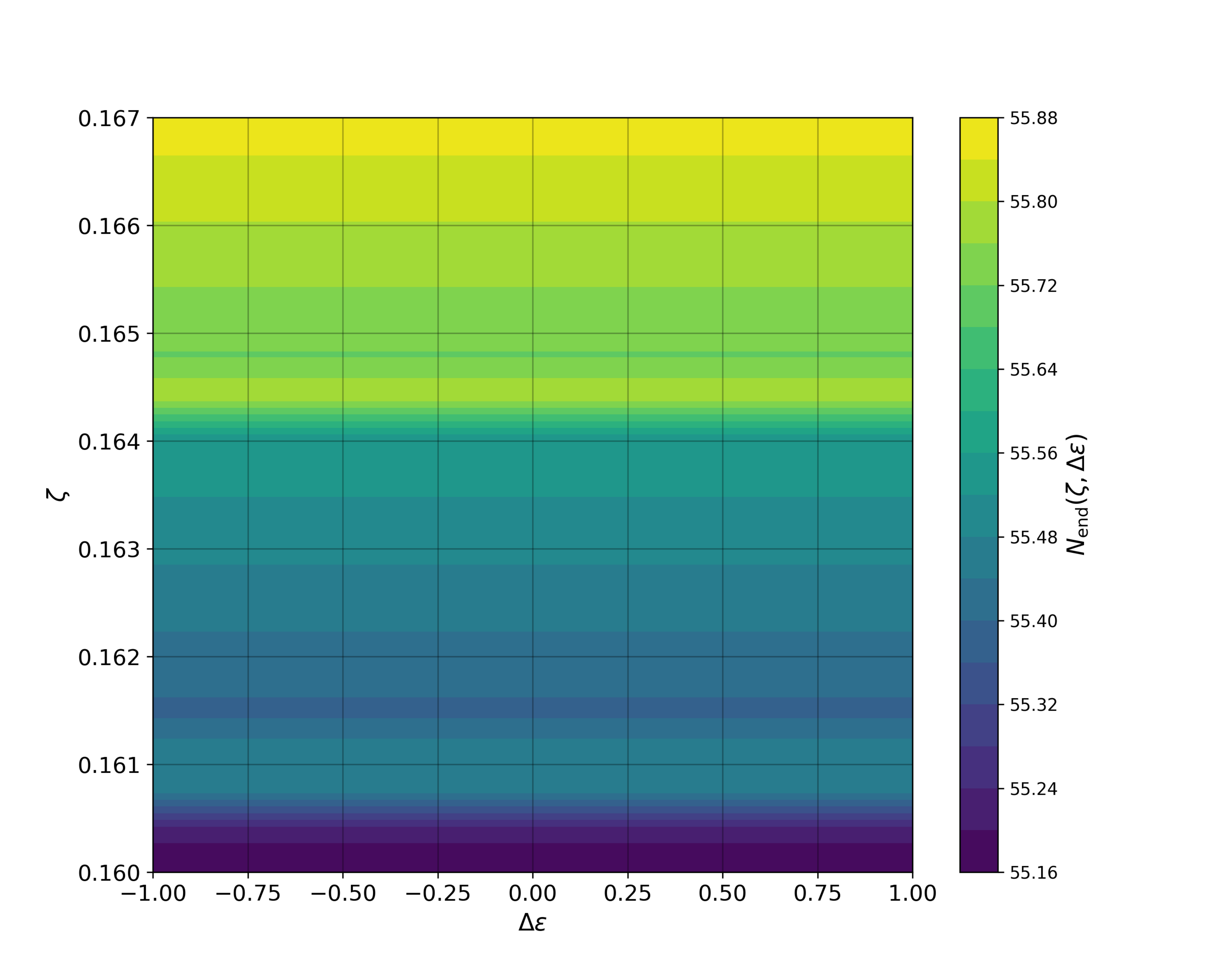}
    \includegraphics[width=0.49\linewidth]{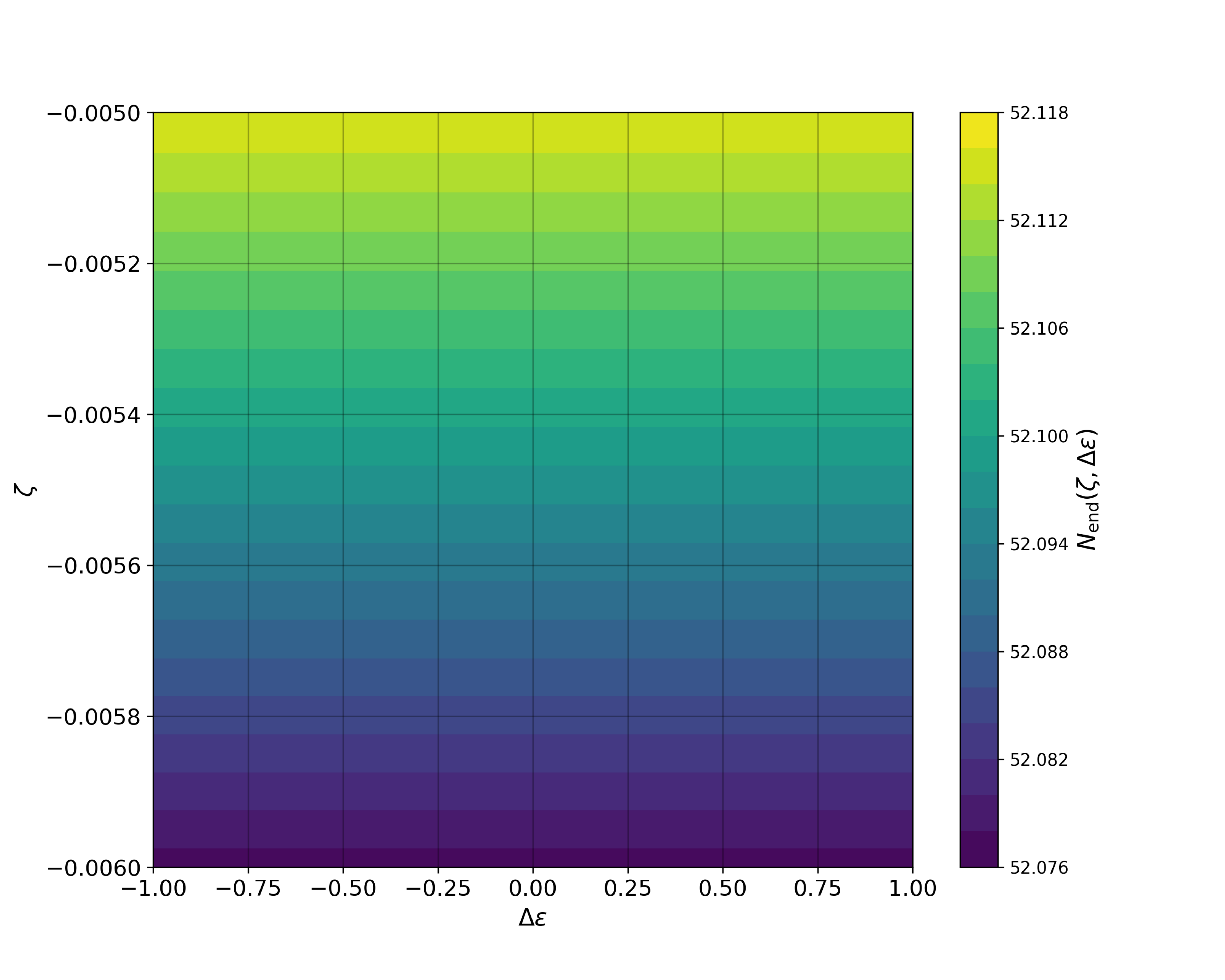}
    \caption{Number of e-folds $N_{\text{end}}(\zeta,\Delta\varepsilon)$ as a function of the nonminimal coupling $\zeta$ and the asymmetry parameter $\Delta\varepsilon$. Left: $\mu = 0.1100$ with $\zeta \in [0.1600,0.1670]$ near the conformal value. Right: $\mu = 0.1000$ with $\zeta \in [-0.0060,0.0050]$. In both panels, $\Delta\varepsilon \in [-1.0,1.0]$. The maps show that $\zeta$ controls the duration of inflation, while $\Delta\varepsilon$ has a negligible impact on the real background evolution.}
    \label{fig:map}
\end{figure}

We first study the influence of the pair $(\zeta,\Delta\varepsilon)$ on the number of e-folds necessary for inflation: $N_{\rm end}(\zeta,\Delta\varepsilon)$. The corresponding results are presented in the two panels of Figure \ref{fig:map}. %(see Appendix \ref{app:appendix1} for details about the construction of these maps). 
For the upper panel, we fix $\mu=0.1100$ and restrict the analysis to $\zeta\in[0.1600,0.1670]$ (near the conformal value). On the other hand, for the lower panel, we use $\mu=0.1000$ and $\zeta\in[-0.0060,0.0050]$. These ranges for $\zeta$ and specific values for $\mu$ were based on the best results for the usual value estimated to $N_{\rm end}\approx 50\sim60$ \cite{liddle.2003,german.2023}. For both panels, we set the wide range for $\Delta\varepsilon\in[-1.0,1.0]$. These maps confirm that $\Delta\varepsilon$ has a negligible influence on the real background expansion, consistent with the expectation that it primarily controls the non-Hermitian sector rather than the global inflationary trajectory. 

\begin{figure}[t!]
    \centering
    \includegraphics[width=0.5\linewidth]{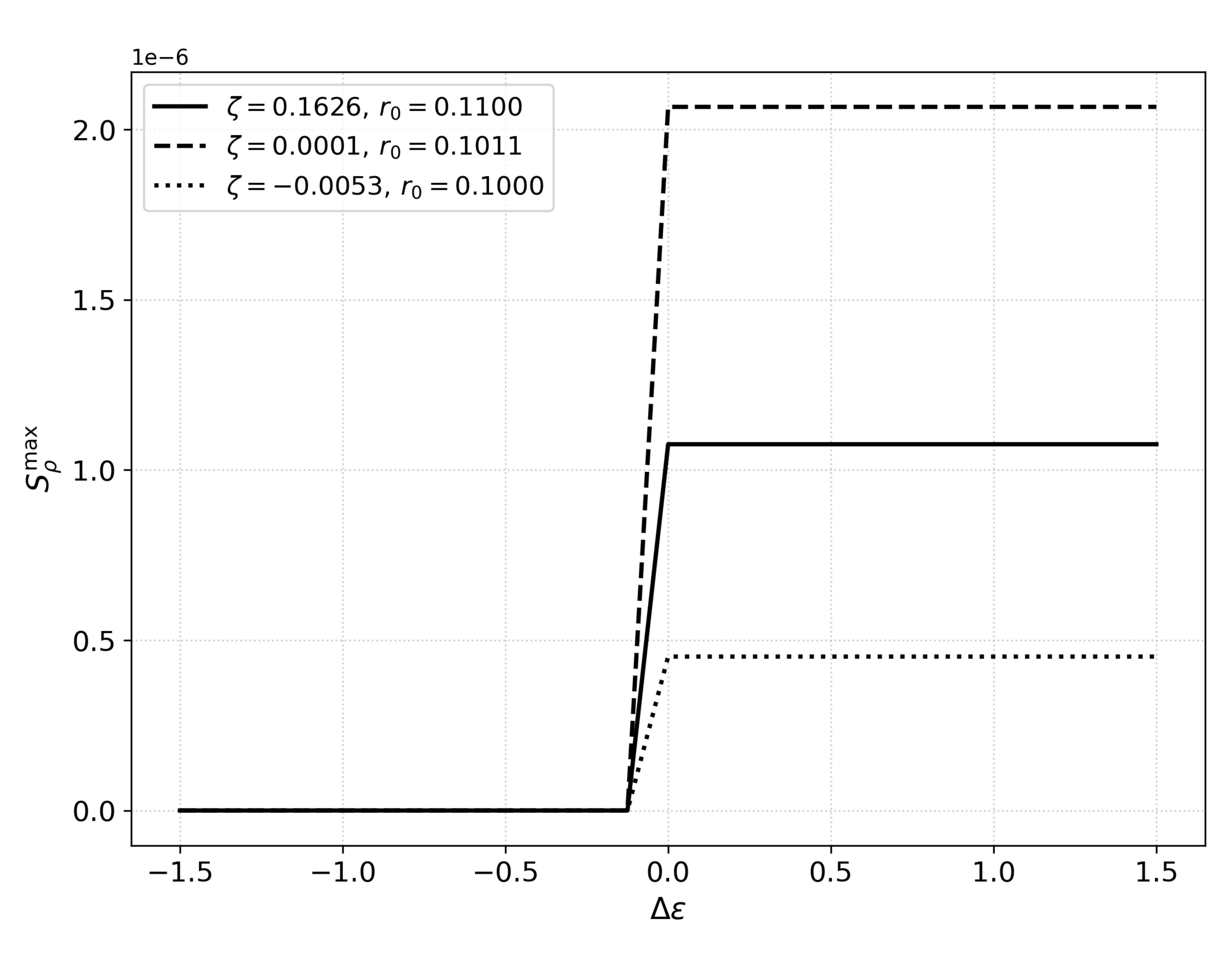}
    \caption{$S_{\rho}^{\max}$ remains extremely small across the scanned parameter space, confirming that the real background evolution is essentially insensitive to the non-Hermitian deformation of the potential.}
    \label{fig:sp}
\end{figure}

To quantify the extent to which the asymmetry parameter $\Delta\varepsilon$ influences the background dynamics, we introduce a sensitivity measure based on finite differences. Specifically, we define
\begin{equation}
\nonumber S_\rho(N) \equiv
\frac{\rho_R(N;\Delta\varepsilon+\delta)-\rho_R(N;\Delta\varepsilon-\delta)}{2\delta},
%\qquad
%S_p(N) \equiv
%\frac{p_R(N;\Delta\varepsilon+\delta)-p_R(N;\Delta\varepsilon-\delta)}{2\delta},
\end{equation}
with $\delta$ chosen sufficiently small to probe the linear response regime. This quantity measures the local sensitivity of the real energy density to variations in $\Delta\varepsilon$, while fully accounting for the backreaction of the complex sector on the field dynamics. As illustrated in Figure \ref{fig:map}, the effect of $\Delta\varepsilon$ on the real part of the energy density is negligible and can therefore be disregarded. Consequently, the imaginary deformation of the potential acts as a perturbative extension: it introduces an effective non-Hermitian, $\mathcal{PT}$-symmetric channel for energy transfer~\cite{bender.1998,bender.2007,mostafazadeh.2010} that remains subdominant during slow-roll but becomes dynamically relevant near the end of inflation \cite{ashida.2020}. Moreover, Figure \ref{fig:sp} shows  
\begin{equation}\label{eq:smax}
S_\rho^{\max} \equiv \max_{N\in[\rm min,\rm max]} |S_\rho(N)|.
\end{equation}

%For the parameter-space analysis presented in the preceding subsection, we compress the time dependence into scalar diagnostics by taking the maximum absolute response over a fixed interval of e-folds,
%\begin{equation}
%S_\rho^{\max} \equiv \max_{N\in[\rm min,\rm max]} |S_\rho(N)|.
%\qquad
%S_p^{\max} \equiv \max_{N\in[\rm min,\rm max]} |S_p(N)| .
%\end{equation} %Based on these findings, we fix $\Delta\varepsilon = 1.0$ for the rest of this analysis and select two representative values for the non-minimal coupling: $(\zeta,\mu) = (0.1626,0.1100)$ and $(\zeta.\mu) = (-0.0053,0.1000)$.

The maximum sensitivity $S_{\rho}^{\max}$ of the real energy density to changes in the asymmetry parameter $\Delta\varepsilon$ remains many orders of magnitude below unity for all the trajectories explored, indicating that the complex deformation acts only as a tiny perturbation on the real FRW background. In other words, the duration of inflation and the overall background dynamics are almost entirely controlled by the real part of the potential and by the nonminimal coupling $\zeta$ (see Figure \ref{fig:map}), whereas the imaginary sector remains effectively decoupled during slow roll.

%================
\subsection{Relevance of the imaginary sector}
\label{sec:reheating}

Since the inflaton is governed by a complex potential $V = V_R + iV_I$, a purely Hermitian comparison with CMB observables is insufficient. We therefore impose a complementary non-Hermitian consistency condition at horizon exit,
\begin{equation}
\nonumber \Delta_{\rm NH} \equiv
\max_{N\in[N_{*}-\Delta N,\,N_{*}+\Delta N]}
\Bigl(|w_I(N)|,\;\mathcal{R}(N)\Bigr),
\label{eq:DeltaNH_eff}
\end{equation}
where $w_I(N)$, $\mathcal{R}(N)$ measure the relative importance of the imaginary sector, and $\Delta N\sim\mathcal{O}(1)$ specifies a window around the $N_*$. Requiring $\Delta_{\rm NH}\ll 1$ ensures that the non-Hermitian deformation is negligible during the quasi–de Sitter phase relevant for the generation of primordial perturbations, while allowing it to become dynamically significant near and after the end of inflation, consistent with its interpretation as an effective decay (reheating) channel. Then, the dissipative behavior becomes important only after the slow-roll regime ends. In this subsequent phase, $|w_I|$ and $\mathcal{R}(N)$ grow rapidly, driving efficient reheating without perturbing the real FRW background during the observable inflationary window.

\begin{figure}
    \centering
\includegraphics[width=0.49\linewidth]{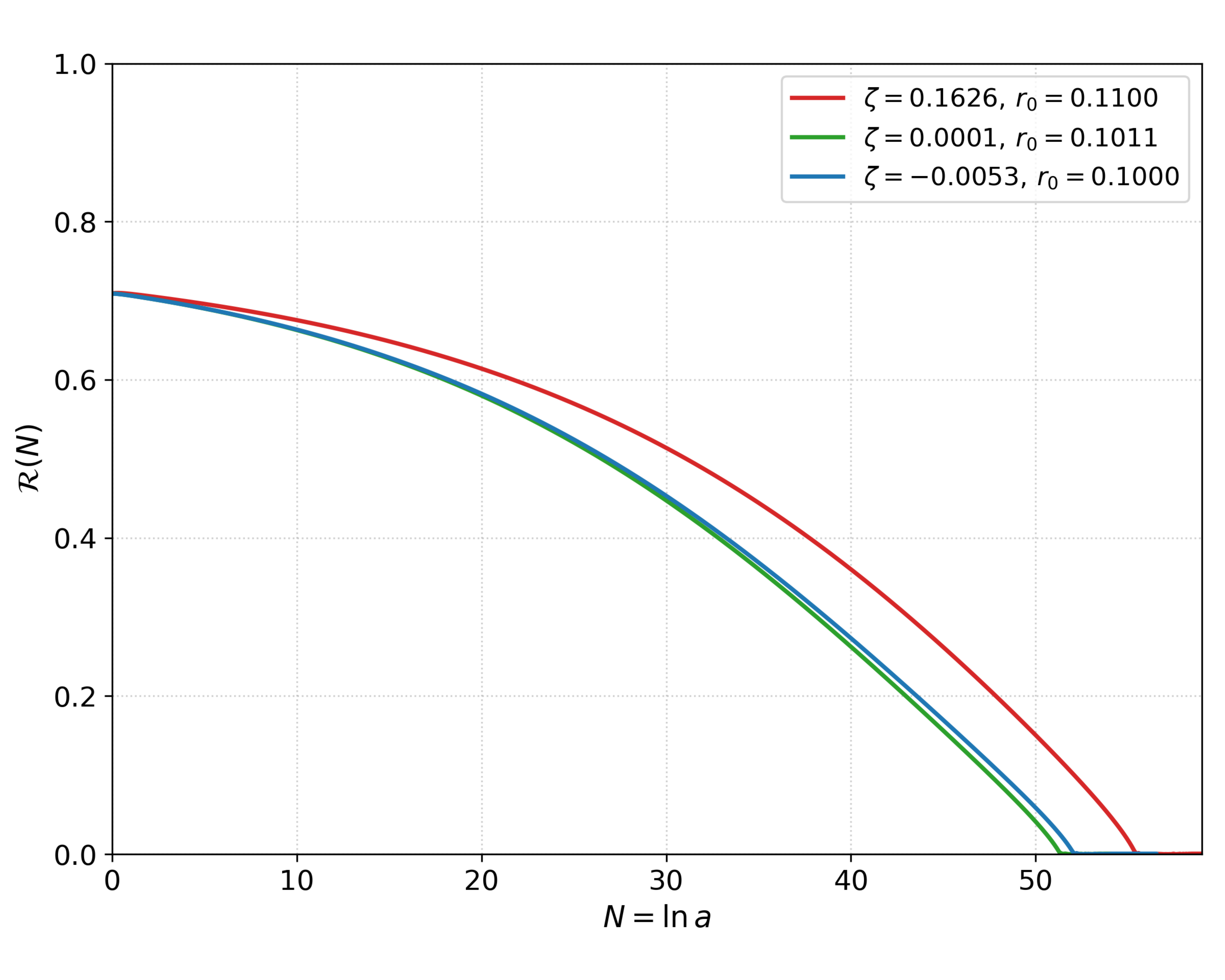}
\includegraphics[width=0.49\linewidth]{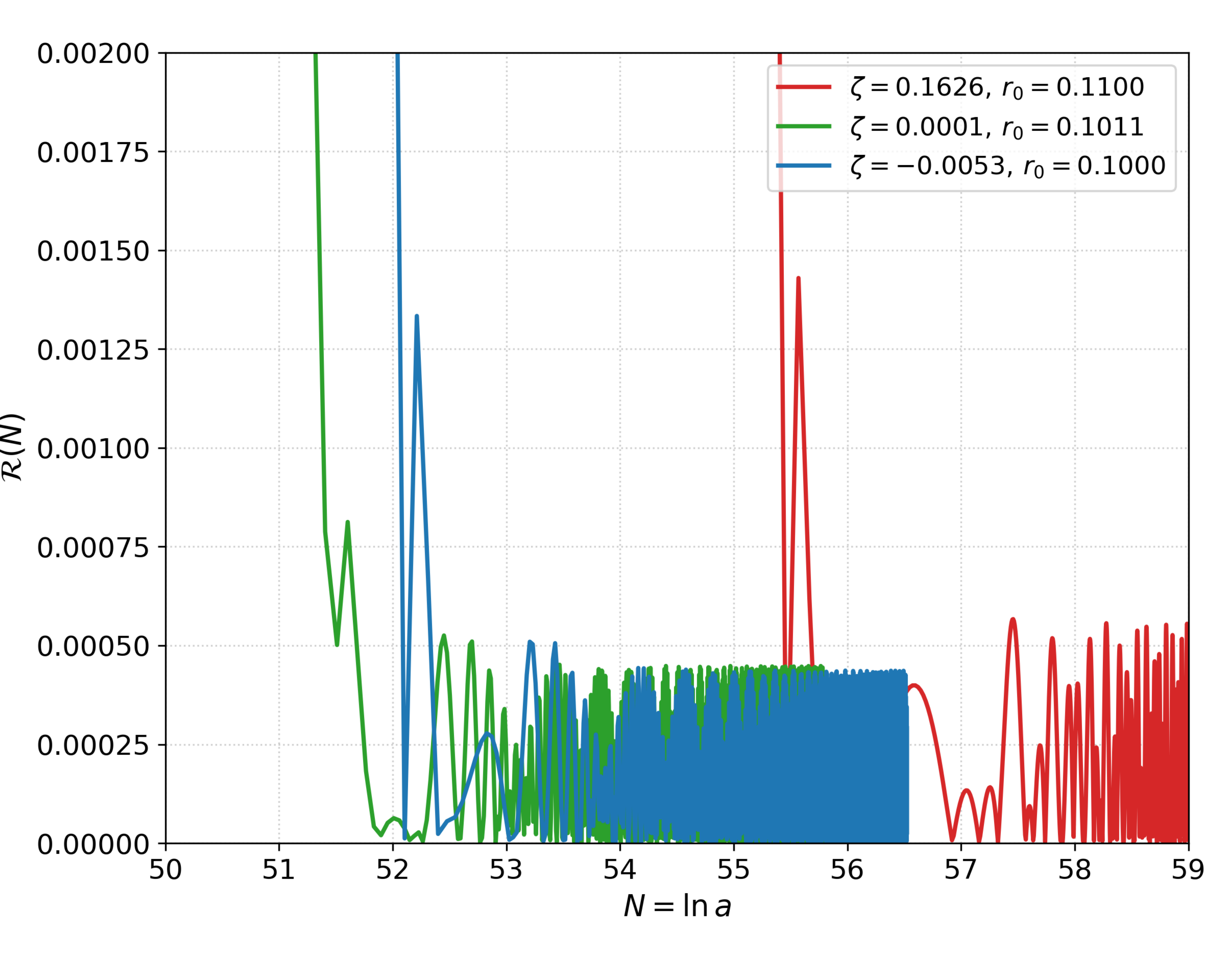}
    \caption{Relevance parameter $\mathcal{R}(N)$ given by equation \eqref{eq:relevance} as a function of the number of e-folds $N = \ln a$ for the parameter choices of Table \ref{tab:comparison}. For small $N$, the imaginary sector contributes a sizeable fraction of the total energy density ($R \sim \mathcal{O}(1)$), but its relevance decreases as inflation proceeds and becomes negligible near and after $N_{\text{end}}$, where reheating takes place.}
    \label{fig:relevance}
\end{figure}

Standard single-field inflation models introduce phenomenological decay terms or friction coefficients to describe reheating, (artificially) transferring energy from the inflaton to other degrees of freedom~\cite{bassett.2006}. In this framework, reheating is a dynamical process arising from a non-Hermitian deformation of the inflaton potential. The role of the non-Hermitian sector is conveniently characterized by the imaginary component of the effective EoS \eqref{eq:w_def}
%\begin{equation}
%\nonumber w_I \equiv \Im\!\left(\frac{p}{\rho}\right)=-\frac{V_I(\rho_R+p_R)}{\rho_R^2+V_I^2},
%\end{equation}
and by the relevance parameter defined as
\begin{equation}\label{eq:relevance}
\mathcal{R}(N)=\frac{|\rho_I|}{\sqrt{\rho_R^{\,2}+\rho_I^{\,2}}}\in[0,1],
\end{equation}
which quantifies the imaginary contribution as a fraction of the real background energy density. Figure \ref{fig:relevance} shows the results for the relevance parameter \eqref{eq:relevance} according to $N$. Observe that for $N\lesssim 10$ and $\rho_R\approx \rho_I$, one has $\mathcal{R}(N)\approx 1/\sqrt{2}$. It is interesting to note that as $N$ increases, the real part of the energy density dominates; however, up to $N\approx 40$, the imaginary part of the energy density still represents nearly 20$\%$ of the total energy density. Near and above $N_{\rm end}$, the relevance of the imaginary sector tends to vanish; this means this sector can be viewed as a perturbative aspect for the real sector.  

\begin{figure}[t!]
    \centering
    \setlength{\fboxsep}{5pt}
    \setlength{\fboxrule}{1pt}
   \includegraphics[width=0.49\linewidth]{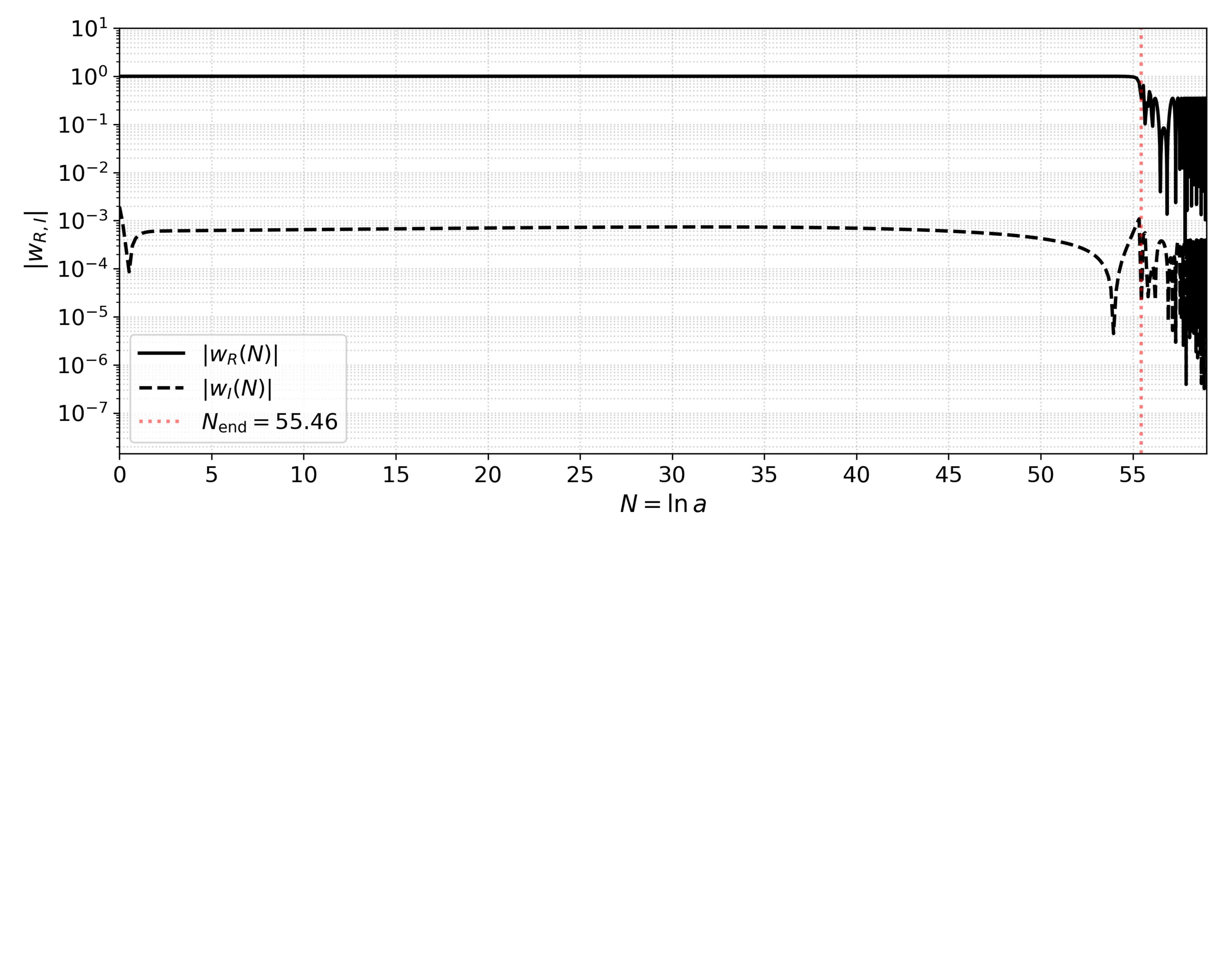}
    \includegraphics[width=0.49\linewidth]{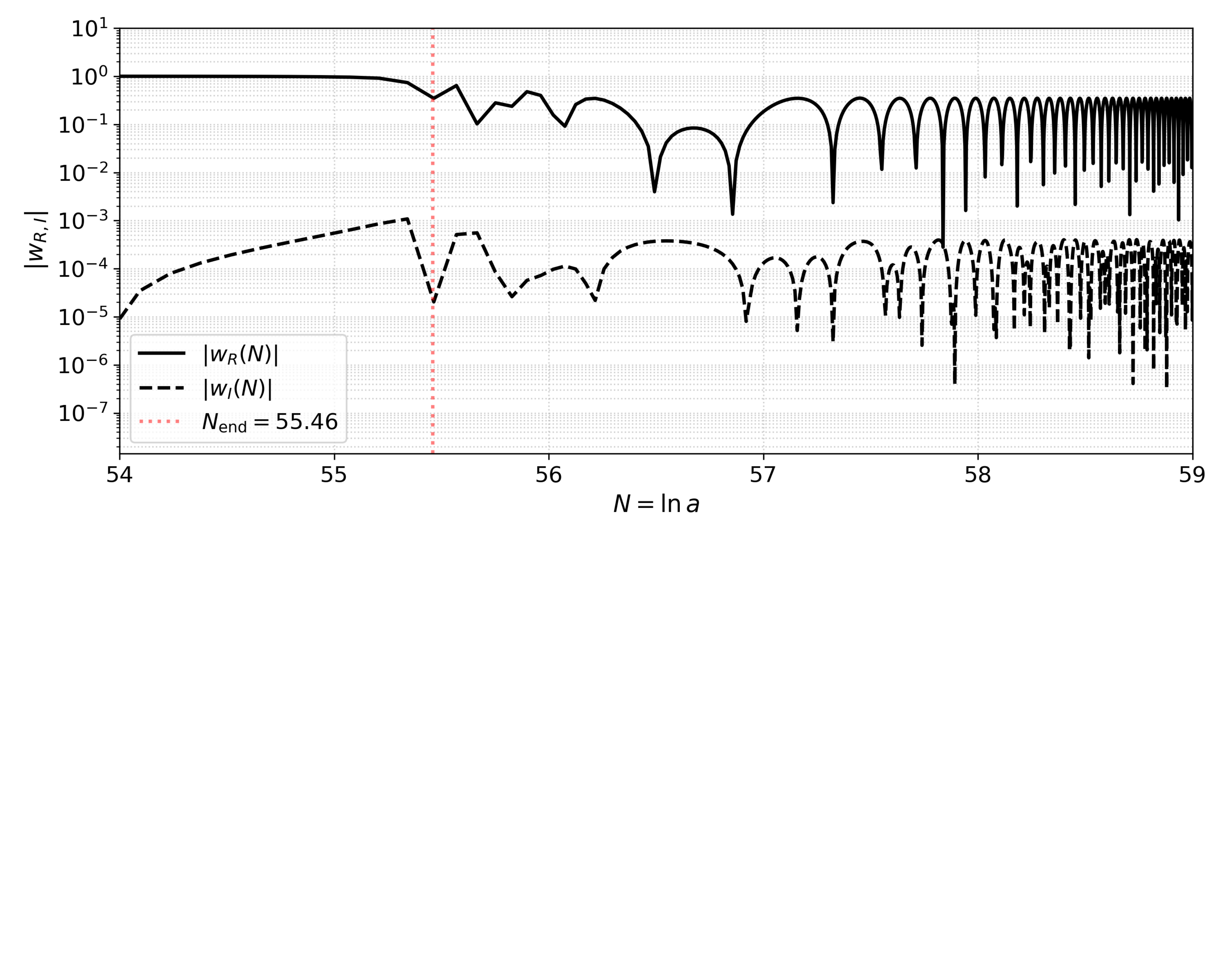}\\ \vspace{-2.7cm}
    \includegraphics[width=0.49\linewidth]{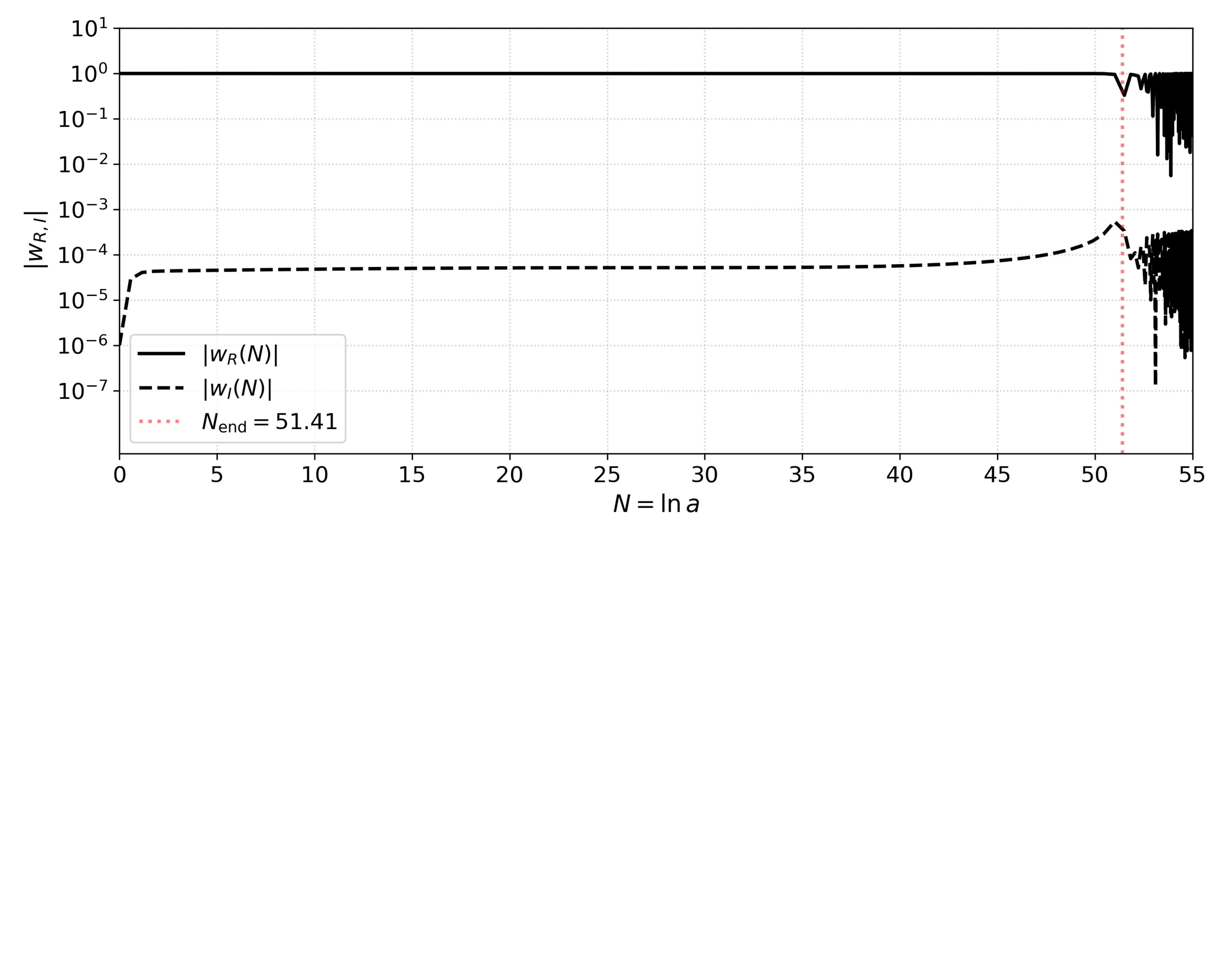}    \includegraphics[width=0.49\linewidth]{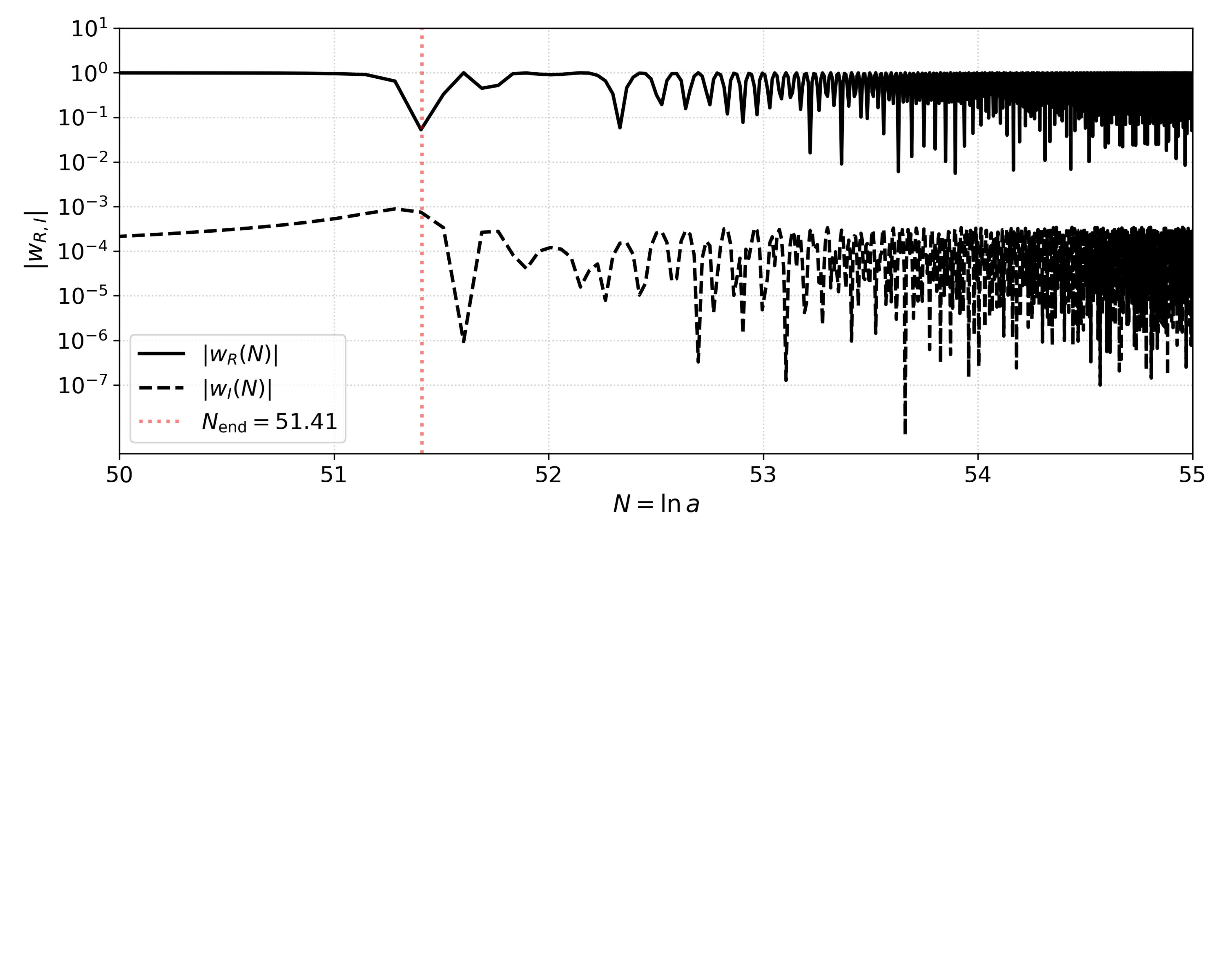}\\ \vspace{-2.7cm}
    \includegraphics[width=0.49\linewidth]{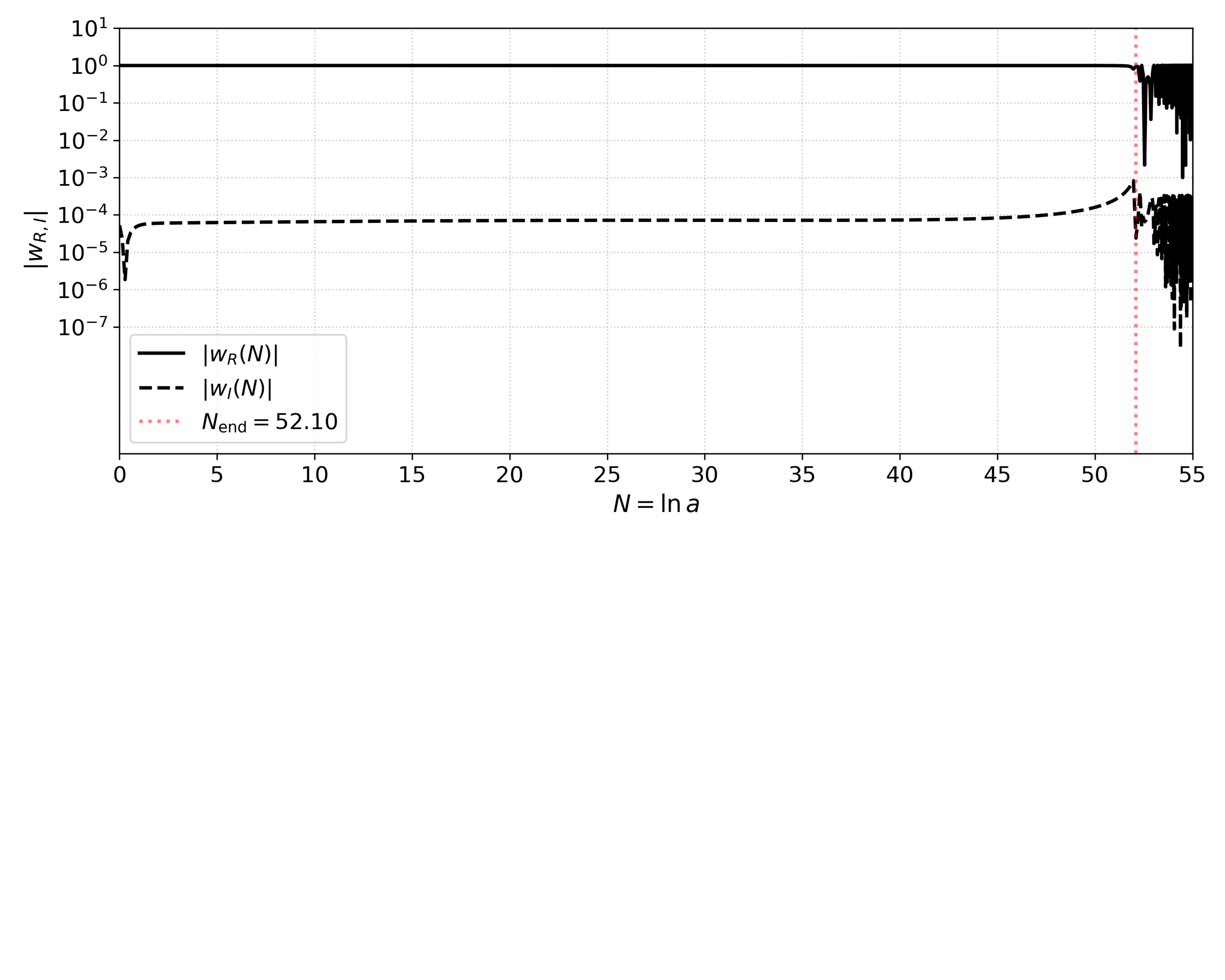}    \includegraphics[width=0.49\linewidth]{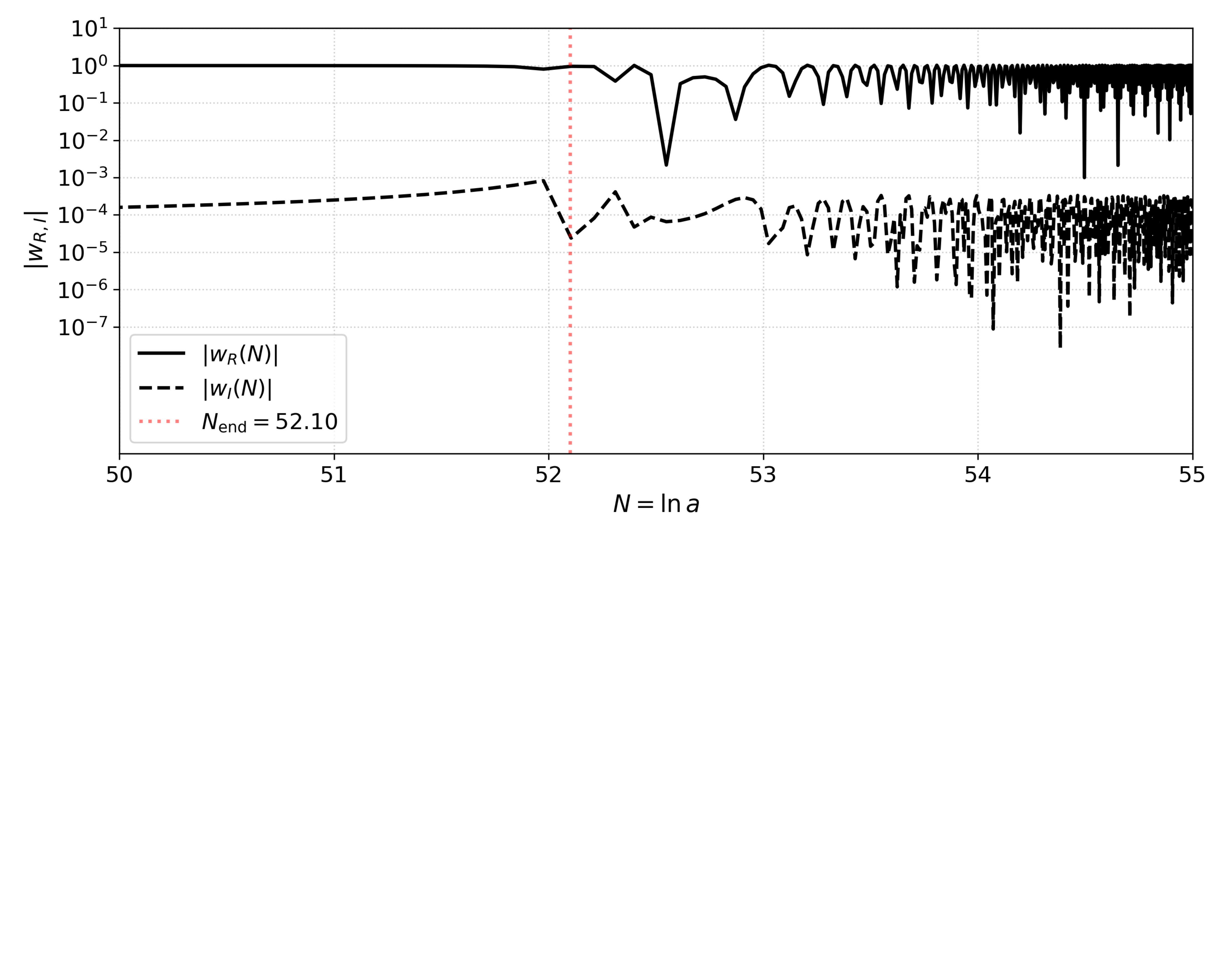}\vspace{-2.7cm}
    \caption{Absolute values of the real and imaginary parts of the effective equation of state, $|w_R(N)|$ and $|w_I(N)|$, as functions of the number of e-folds $N = \ln a$. The plots illustrate that during slow roll the Universe undergoes a quasi--de Sitter phase with $w_R \simeq -1$ and $|w_I| \ll 1$, while near the end of inflation $|w_I|$ grows and transiently enhances dissipation, initiating the reheating phase.}
\label{fig:w_evolution}
\end{figure}

%%%%%%%%%%%%%%%%%%%%%%%%%%%%%%%%%%%%5
\subsection{Real and Imaginary parts of $w(N)$} 

%We start by defining the comoving horizon crossing as $k=aH$.
Taking into account the best values for ($\zeta,\mu)$ and fixed $\Delta\varepsilon=1.0$, Figure \ref{fig:w_evolution} displays the evolution of $w_R$ and $w_I$ as functions of the e-fold number $N$ for two representative pairs: for the left panel, $(\zeta,\mu) = (0.1626,0.1100)$, and for the right panel, $(\zeta,\mu) = (-0.0053,0.1000)$. Within the initial conditions, the inflaton starts high on the plateau of $V_R$, enabling efficient slow-roll that lasts approximately 55 e-folds ($\zeta = 0.1626$) or 52 e-folds ($\zeta =-0.0053$) before the potential slope and non-minimal term drive $w_R$ away from $-1$~\cite{dimarco.2024}. Throughout the slow-roll regime, $V_R$ dominates, and the asymmetry term $V_I = \Delta\varepsilon\,\phi\chi$ remains small, yielding $|w_I| \ll 1$. This confirms that non-Hermitian effects are negligible during inflation and only become significant near its end. %The real part $w_R \approx -1$ governs the background expansion, while $w_I$ quantifies the non-Hermitian contribution generated by the imaginary potential component $V_I$.

\subsection{Connection with non-Hermitian field theory and open systems}

The complex-valued scalar potential places this model within the broader setting of non-Hermitian field theories and effective open-system descriptions \cite{rotter.2009}. %In these frameworks, complex terms in the action do not indicate inconsistency but represent irreversible processes such as decay, dissipation, or coarse-grained interactions with unobserved degrees of freedom. 
Established examples include complex mass poles of unstable particles in QFT and effective Hamiltonians in condensed matter with finite decay widths \cite{Bogoliubov1959,Rothe2006,HatanoNelson1996}.

In the present cosmological setting, the imaginary part of the potential, $V_I = \Delta\varepsilon\,\phi\chi$, plays a similar role. Although this modification renders the scalar sector formally non-Hermitian, its physical effects are fully encoded in the coupled equations of motion, where they emerge dynamically as a channel for energy transfer. By sourcing the imaginary components of the fields, $V_I$ introduces a controlled departure from conservative dynamics, modeling dissipation without requiring a complex spacetime metric.

The prescription used here—letting the real part of the energy-momentum tensor determine spacetime geometry while interpreting the imaginary part as a source of dissipation—matches standard views in non-Hermitian quantum mechanics \cite{Mostafazadeh2002,BreuerPetruccione}, where the real part of the spectrum typically governs coherent evolution, and the imaginary part determines the lifetimes and decay rates of spectrum sets. Similar effective approaches have been applied in cosmology to model particle production, reheating, and out-of-equilibrium damping of scalar fields \cite{CalzettaHu,Boyanovsky1996}.

%In this interpretation, the sensitivity measures $S_\rho^{\max}$ and $S_p^{\max}$ quantify the backreaction of the non-Hermitian sector on the real background dynamics. Non-vanishing sensitivities mean that the complex sector does more than provide dissipation: it dynamically modifies the real energy density and pressure. Vanishing sensitivities instead correspond to a regime where the imaginary potential causes dissipation without significantly affecting the Friedmann evolution. Thus, the framework offers a continuous, controlled interpolation between conservative inflationary dynamics and effective open-system behavior.

%%%%%%%%%%%%%%%%%%%
\subsection{Relation to $\mathcal{PT}$-symmetric field theories}

Non-Hermitian quantum theories with $\mathcal{PT}$ symmetry form a well-established class of models where real spectra and stable dynamics can occur without standard Hermiticity \cite{Bender1998,bender.2007}. Here, the combined spatial parity $\mathcal{P}$ and time-reversal $\mathcal{T}$ symmetries balance imaginary terms in the Hamiltonian, keeping physical observables real in the unbroken symmetry phase.

In this model, the imaginary potential $V_I = \Delta\varepsilon\,\phi\chi$ serves as an effective $\mathcal{PT}$-symmetric deformation\footnote{One can always choose $\phi$ and $\chi$ with opposite intrinsic parities so that, under $\mathcal{P}$, the product $\phi\chi$ changes sign, while the factor $i$ originating from the imaginary sector changes sign under $\mathcal{T}$, so that the combination $i\Delta\varepsilon\phi\chi$ remains invariant under $\mathcal{PT}$.}. With suitable intrinsic parity assignments for the scalar fields, the total potential $V = V_R + iV_I$ remains invariant under $\mathcal{PT}$. %Thus, the non-Hermitian term from $\Delta\varepsilon$ is not pathological but places the system within the broader class of $\mathcal{PT}$-symmetric effective field theories.

From this perspective, the imaginary sector plays the role of gain-loss terms in $\mathcal{PT}$-symmetric optical or condensed-matter systems, mediating energy exchange between coupled modes. In our cosmological setting, this appears as a transfer between the scalar sector and the gravitational background, supporting the interpretation of $V_I$ as a decay or reheating channel \cite{ashida.2020}.

\section{Remarks and Conclusions}\label{sec:final}

We have analyzed an inflationary model driven by a complex scalar field with a non-symmetric potential and non-minimal gravitational coupling. This setup generates a non-Hermitian dynamical system, which we treat within the framework of EFT. By decomposing the dynamics into a real sector that determines the spacetime background and an imaginary sector that encodes dissipative and decay processes, we have formulated a theoretical framework for describing both the inflationary phase and its end.

Concerning the questions in the Introduction of this work, the introduction of the CIF furnishes the following answers: (i) complex potentials do sustain viable inflationary backgrounds; (ii) imaginary effects are suppressed during slow roll but grow dramatically near the exit from inflation; (iii) the inflationary observables $n_s$ and $r$ are insensitive to $\Delta\varepsilon$, with sensitivity variations of order $10^{-5}$, confirming that the real background is decoupled from the non-Hermitian deformation; (iv) effective reheating emerges from the imaginary sector without extra fields; and (v) $\zeta$ is the primary control parameter for tuning the background, while $\Delta\varepsilon$ controls the strength of dissipation.

This analysis shows that the inflationary background is robust against non-Hermitian deformations. Our sensitivity study shows that the energy density and pressure are insensitive to the asymmetry parameter $\Delta\varepsilon$ during slow roll. Thus, the duration of inflation and the expansion history depend almost entirely on the real part of the potential and the non-minimal coupling $\zeta$, while the imaginary sector behaves as a spectator field until inflation ends.

For non-minimal couplings $\zeta \sim 0.06$–$0.14$, the spectral index $n_s$ and tensor-to-scalar ratio $r$ lie within the $1\sigma$–$2\sigma$ Planck 2018 contours. These observables are independent of $\Delta\varepsilon$, so trajectories with different non-Hermitian strengths yield a single prediction curve in the $n_s\!-\!r$ plane. This decoupling lets the model meet CMB constraints while still allowing non-conservative behavior at the end of inflation.

The reheating mechanism arises geometrically from the imaginary sector. The relevance parameter $\mathcal{R}(N)$ remains negligible ($\mathcal{R}(N) \ll 1$) during the quasi-de Sitter phase, then rapidly grows to $\mathcal{O}(1)$ right after inflation, showing chaotic oscillations. This effectively reproduces the phenomenological friction term ($\Gamma \dot{\phi}$) of standard single-field models, but here it follows from the complex potential’s structure. The imaginary sector acts as a heat bath, although a full microphysical account of reheating still requires coupling it to Standard Model fields to describe the final thermalization of the condensate’s released energy.

Complex or negative squared masses in this framework do not indicate pathology or superluminal propagation. Viewed through $\mathcal{PT}$-symmetric and open quantum systems, they encode vacuum instabilities and the open-system nature of the inflaton in an expanding spacetime. This perspective suggests new research directions, including preheating with non-Hermitian couplings and a possible non-Gaussian signal from the complex phase. We conclude that properly defined complex inflaton potentials provide a minimal, geometrically unified origin for both accelerated expansion and the subsequent reheating of the Universe.

\section*{Acknowledgments}

SDC acknowledges the Federal University of São Carlos (UFSCar) and the Applied Mathematics Laboratory (CCTS/DFQM) for their institutional support.


\begin{thebibliography}{99}
%1
\bibitem{riess.1998}A. G. Riess {\it et al.}: {\it  Observational evidence from supernova for an accelerating universe and cosmological constant}. Astron. J. 116, 1009–1038 (1998).

\bibitem{perlmutter.1999}S. Perlmutter {\it et al.}: {\it  Measurements of omega and lambda from 42 high-redshift supernovae}. Astrophys. J. 517, 565–585 (1999).

\bibitem{jaman_sami.2022}N. Jaman and M. Sami: {\it  What is needed of a scalar field if it is to unify inflation and late time acceleration?} Galaxies 10, 51 (2022).

\bibitem{berera.1995}A. Berera: {\it  Warm inflation}. Phys. Rev. Lett. 75, 3218 (1995).

\bibitem{kinney.2005}W. H. Kinney: {\it  Horizon crossing and inflations with large $\eta$}. Phys. Rev. D 72, 023515 (2005).

\bibitem{cheung.2008}C. Cheung {\it et al.}: {\it  The effective field theory of inflation}. JHEP 03, 014 (2008).

\bibitem{kobayashi.2011}T. Kobayashi, M. Yamaguchi, and J. Yokoyama: {\it  Generalized $G$-inflation}. Prog. Theor. Phys. 126, 511 (2011).

\bibitem{martin.2014}J. Martin, C. Ringeval, and V. Vennin: {\it  Encyclopaedia inflationaris}. Phys. Dark Univ. 5-6, 75–235 (2014).

\bibitem{baumann.2015}D. Baumann and L. McAllister: {\it  Inflation and string theory}. Cambridge Univ. Press, (2015).

\bibitem{pinol.2015}L. Pinol and S. Renaux-Petel: {\it  The effective field theory of inflation beyond slow-roll}. Phys. Rev. D 92, 083513 (2015).

\bibitem{achucarro.2018}A. Achúcarro {\it et al.}: {\it  Effective theories of single inflation when heavy fields matter}. JHEP 01, 131 (2018).

\bibitem{cabass.2021}G. Cabass, E. Pajer, and D. van der Woude: {\it  The effective field theory of inflationary perturbations}. JHEP 08, 050 (2021).

\bibitem{yurov.2002}A. V. Yurov: {\it  Complex field as inflaton and quintessence}. arXiv:hep-th/0208129v1.

\bibitem{dias.frazer.2016}M. Dias, J. Frazer, and M. C. David Marsh: {\it  Simple emergent power spectra from complex inflationary physics}. Phys. Rev. Lett. 117, 141303 (2016).

\bibitem{zhang.zheng.2022}R. Zhang and S. Zheng: {\it  Complex inflaton field}. arXiv:2106.00977.

\bibitem{basterogil.2021}M. Bastero-Gil {\it et al.}: {\it  Towards a reliable effective field theory of inflation}. Phys. Lett. B 813, 136055 (2021).

\bibitem{denner.2020}A. Denner and S. Dittmaier: {\it  The complex-mass scheme for perturbative calculations with unstable particles}. Phys. Rept. 864, 1–163 (2020).

\bibitem{ashida.2020}Y. Ashida, Z. Gong, and M. Ueda: {\it  Non-hermitian physics}. Adv. Phys. 69, 249 (2020).

\bibitem{chavanis.2022}P.-H. Chavanis: {\it  Cosmological models based on a complex scalar field with a power-law potential associated with a polytropic equation of state}. Phys. Rev. D 106, 043502 (2022).

\bibitem{ahmad.2019}S. Ahmad {\it et al.}: {\it  Baryogenesis in the paradigm of quintessential inflation}. Phys. Rev. D 100, 103525 (2019).

\bibitem{khalatnikov.1992}I. M. Khalatnikov and A. Mezhlumian: {\it  The classical and quantum cosmology with a complex scalar field}. Phys. Lett. A 169, 308–312 (1992).

\bibitem{khalatnikiv.1994}L. Amendola, I. M. Khalatnikov, M. Litterio, and F. Occhionero: {\it  Quantum cosmology with a complex field}. Phys. Rev. D 49, 1881 (1994).

\bibitem{bertolami.2008}O. Bertolami, F. S. N. Lobo, and J. Páramos: {\it  Nonminimal coupling of perfect fluids to curvature}. Phys. Rev. D 78, 064036 (2008).

\bibitem{faraoni.2004}V. Faraoni: {\it  Cosmology in scalar-tensor gravity}. Kluwer Academic Publ., (2004).

\bibitem{I.M.Khalatnikov.1997}A. Yu. Kamenshchik, I. M. Khalatnikov, and A. V. Toporensky: {\it  Complex inflaton field in quantum cosmology}. Int. J. Mod. Phys. D 6, 649–672 (1997).

\bibitem{burgess.2007}C. P. Burgess: {\it  Introduction to effective field theory}. Ann. Rev. Nucl. Part. Sci 57, 329 (2007).

\bibitem{kallosh.2013_1}R. Kallosh, A. Linde, and D. Roest: {\it  Superconformal inflationary $\alpha$-attractors}. JHEP 2013, 198 (2013).

\bibitem{kallosh_2013_2}R. Kallosh and A. Linde: {\it Universality class in conformal inflation}. JCAP 07, 002 (2013).

\bibitem{kallosh.2014}R. Kallosh, A. Linde, and D. Roest: {\it Large field inflation and double $\alpha$-attractors}. JHEP 2014, 052 (2014).

\bibitem{dimopoulos.2018}K. Dimopoulos and C. Owen: {\it Quintessential inflation with $\alpha$-attractors}. arXiv:1703.00305v3.

\bibitem{shojaee.2021}R. Shojaeea, K. Nozarib, and F. Darabia: {\it $\alpha$-attractors and reheating in a non-minimal inflationary model}. Int. J. Mod. Phys. D 29(10), 2050077 (2020).

\bibitem{mishra.2025}S. S. Mishra and V. Sahni: {\it New models of quintessential inflation featuring plateau and hilltop potentials}. Eur. Phys. J. C 85, 48 (2025).

\bibitem{bhattacharya.2023}S. Bhattacharya, K. Dutta, M. R. Gangopadhyay, and A. Maharana: {\it $\alpha$-attractor inflation: models and predictions}. Phys. Rev. D 107(10), 103530 (2023).

\bibitem{baumann.2012}D. Baumann: {\it Tasi lectures on inflation}. arXiv:0907.5424v2.

\bibitem{stein.2021}N. K. Stein and W. H. Kinney: {\it Natural inflation after Planck 2018}. arXiv:2106.02089v2.

\bibitem{planck.collab.2015}N. Aghanim {\it et al.} (Planck Collaboration): {\it Planck 2018 results. VI. Cosmological parameters}. A$\&$A 641, A6 (2020).

\bibitem{linde.2025}A. Linde: {\it Alexei Starobinsky and modern cosmology}. arXiv:2509.01675v2.

\bibitem{pozo.2024}D. Pozo {\it et al.}: {\it Some inflationary models under the light of Planck 2018 results}. arXiv:2311.04683.

\bibitem{cook.2015}J. L. Cook {\it et al.}: {\it Reheating predictions in single field inflation}. arXiv:1502.04673.

\bibitem{gorini.2004}V. Gorini, A. Kamenshchik, U. Moschella, and V. Pasquier: {\it Tachyons, scalar fields, and cosmology}. Phys. Rev. D 69, 123512 (2004).

\bibitem{wald.1984}R. Wald: {\it General Relativity}. University of Chicago Press, (1984).
17

\bibitem{ford.1987}L. H. Ford: {\it Gravitational particle creation and inflation}. Phys. Rev. D 35(10), 2955 (1987).

\bibitem{salopek.1989}D. S. Salopek, J. R. Bond, and J. M. Bardeen: {\it Designing density fluctuation spectra in inflation}. Phys. Rev. D 40(6), 1753 (1989).

\bibitem{tsujikawa.2000}S. Tsujikawa: {\it Power-law inflation with a nonminimally coupled scalar field}. Phys. Rev. D 62, 043512 (2000).

\bibitem{nielsen.1978}H. B. Nielsen: {\it Tachyons, monopoles and related topics} (chapter Tachyons in field theory), pages 169–174. E. Recami - North Holland Publ., (1978).

\bibitem{sen.2002}A. Sen: {\it Field theory of tachyon matter}. Mod. Phys. Lett. A 17, 1797–1804 (2002).

\bibitem{padmanabhan.2002}T. Padmanabhan: {\it Accelerated expansion of the universe driven by tachyonic matter}. Phys. Rev. D 66, 021301 (2002).

\bibitem{S.Willenbrock.2024}S. Willenbrock: {\it Mass and width of an unstable particle}. arXiv:2203.11056.

\bibitem{sergeenko.2014}M. N. Sergeenko: {\it Complex masses of resonances in the potential approach}. Nonlinear Phen. in Complex Syst. 17(4), 433–438 (2014).

\bibitem{klein.roest.2016}R. Klein and D. Roest: {\it Exorcising the Ostrogradsky ghost in coupled systems}. J. High Energ. Phys. 130 (2016).

\bibitem{ostrogradsky.1850}M. Ostrogradsky: {\it Mémoires sur les équations différentielles, relatives au problème des isopérimètres}. Mem. Ac. St. Petersbourg, VI(4), 385–517 (1850).

\bibitem{ludwick.2017}K. J. Ludwick: {\it The viability of phantom dark energy: A review}. Mod. Rev. Lett. A 32(28), 1730025 (2017).

\bibitem{vikman.2005}A. Vikman: {\it Can dark energy evolve to the phantom?} Phys. Rev. D 71, 023515 (2005).

\bibitem{singh.2003}P. Singh, M. Sami, and N. Dadhich: {\it Cosmological dynamics of a phantom field}. Phys. Rev. D 68(2), 023522 (2003).

\bibitem{lobo.2025}F. S. N. Lobo, T. Harko, and M. A. S. Pinto: {\it Modified gravity with nonminimal curvature–matter couplings: a framework for gravitationally induced particle creation}. Universe 11(11), 356 (2025).

\bibitem{liddle.2003}A. R. Liddle and S. M. Leach: {\it How long before the end of inflation were observable perturbations produced?} Phys. Rev. D 68, 103503 (2003).

\bibitem{german.2023}G. Germán, R. G. Quaglia, and A. M. M. Colorado: {\it Model independent bounds for the number of e-folds during the evolution of the universe}. JCAP 03, 004 (2023).

\bibitem{bender.1998}C. M. Bender and S. Boettcher, {\it Real spectra in non-hermitian Hamiltonians having $\mathcal{PT}$-symmetry}. Phys. Rev. Lett. 80, 5243–5246 (1998).

\bibitem{bender.2007}C. M. Bender: {\it Making sense of non-hermitian Hamiltonians}. Rep. Prog. Phys. 70, 947 (2007).

\bibitem{mostafazadeh.2010}A. Mostafazadeh: {\it Pseudo-hermitian representations of quantum mechanics}. Int. J. Geom. Meth. Mod. Phys. 7, 1191 (2010).

\bibitem{bassett.2006}B. A. Bassett, S. Tsujikawa, and D. Wands: {\it Inflation dynamics and reheating}. Rev. Mod. Phys. 78, 537 (2006).

\bibitem{dimarco.2024}A. Di Marco, E. Orazi, and G. Pradisi: {\it Introduction to the number of e-folds in slow-roll inflation}. Universe 10(7), 284 (2024).

\bibitem{rotter.2009}I. Rotter: {\it A non-hermitian Hamiltonian operator and the physics of open quantum systems}. J. Phys. A 42, 153001 (2009).

\bibitem{Bogoliubov1959}N. N. Bogoliubov and D. V. Shirkov: {\it Introduction to the theory of quantized fields}. Interscience, (1959).

\bibitem{Rothe2006}H. J. Rothe: {\it Lattice gauge theories: an introduction}. World Scientific, 2006.

\bibitem{HatanoNelson1996}N. Hatano and D. R. Nelson: {\it Localization transitions in non-hermitian quantum mechanics}. Phys. Rev. Lett. 77, 570–573 (1996).

\bibitem{Mostafazadeh2002}A. Mostafazadeh: {\it Pseudo-hermiticity versus $\mathcal{PT}$ -symmetry}. J. Math. Phys. 43, 205–214 (2002).

\bibitem{BreuerPetruccione}H.-P. Breuer and F. Petruccione: {\it The theory of open quantum systems}. Oxford University Press, (2002).

\bibitem{CalzettaHu}E. A. Calzetta and B. L. Hu: {\it Nonequilibrium quantum fields}. Phys. Rev. D 37, 2878–2900 (1988).

\bibitem{Boyanovsky1996}D. Boyanovsky, H. J. de Vega, R. Holman, D.-S. Lee, and A. Singh: {\it Dissipation via particle production in scalar field theories}. Phys. Rev. D 51, 4419–4444 (1995).

\bibitem{Bender1998}C. M. Bender and S. Boettcher: {\it Real spectra in non-hermitian Hamiltonians having $\mathcal{PT}$-symmetry}. Phys. Rev. Lett.80, 5243–5246 (1998).


\end{thebibliography}
\end{document}